\documentclass[%
 reprint,
superscriptaddress,
preprintnumbers,
 amsmath,amssymb,
 aps,
]{revtex4-2}

\usepackage{graphicx}% Include figure files
\usepackage{dcolumn}% Align table columns on decimal point
\usepackage{bm}% bold math
\usepackage{physics}
\usepackage{xcolor}
\usepackage{comment}
\newcommand{\bp}{\mathbf{p}}
\newcommand{\bq}{\mathbf{q}}
\newcommand{\bk}{\mathbf{k}}
\newcommand{\md}{\mathrm{d}}
\begin{document}

\preprint{RESCEU-4/21}

\title{Why Must Primordial Non-Gaussianity Be Very Small?}% Force line breaks with \\

\author{Jason Kristiano}
\email{jkristiano@resceu.s.u-tokyo.ac.jp}
\affiliation{Research Center for the Early Universe (RESCEU), Graduate School of Science, The University of Tokyo, Tokyo 113-0033, Japan}
\affiliation{Department of Physics, Graduate School of Science, The University of Tokyo, Tokyo 113-0033, Japan}

\author{Jun'ichi Yokoyama}
\email{yokoyama@resceu.s.u-tokyo.ac.jp}
\affiliation{Research Center for the Early Universe (RESCEU), Graduate School of Science, The University of Tokyo, Tokyo 113-0033, Japan}
\affiliation{Department of Physics, Graduate School of Science, The University of Tokyo, Tokyo 113-0033, Japan}
\affiliation{Kavli Institute for the Physics and Mathematics of the Universe (Kavli IPMU), WPI, UTIAS, The University of Tokyo, Kashiwa, Chiba 277-8568, Japan}
\affiliation{Trans-scale Quantum Science Institute, The University of Tokyo, Tokyo 113-0033, Japan}


\date{\today}

\begin{abstract}
One-loop correction to the power spectrum in generic single-field inflation is calculated by using standard perturbation theory. Because of the enhancement inversely proportional to the observed red tilt of the spectral index of curvature perturbation, the correction turns out to be much larger than previously anticipated. As a result, the primordial non-Gaussianity must be much smaller than the current observational bound in order to warrant the validity of cosmological perturbation theory.
\end{abstract}

\maketitle

Inflation in the early Universe \cite{Starobinsky:1980te, Sato:1980yn, Guth:1980zm} (see, e.g., \cite{Sato:2015dga} for a review) has been a part of standard cosmology not only to solve the horizon, flatness, and monopole problems, but also to account for  the origin of large-scale structures. Inflation is well described by a homogeneous scalar field dubbed as inflaton $\phi$ in quasi-de Sitter (dS) space. Properties of the inflaton, such as the forms of its kinetic and potential terms as well as its coupling to gravity are yet to be clarified both from theoretical and observational viewpoints.  Quantum fluctuations of scalar field and gravitational field generated during inflation \cite{Starobinsky:1979ty,Mukhanov:1981xt,Starobinsky:1982ee,Hawking:1982cz,Guth:1982ec} serve as probes of its physics that can be tested by observations of cosmic microwave background (CMB) and large-scale structures. 

The simplest class of inflation models is the potential-driven canonical slow-roll inflation where the kinetic term of the inflaton has the canonical form and its potential energy drives inflation. In this model, fluctuations can be expressed approximately as a massless free scalar field in dS space minimally coupled to gravity.  As a result, its linear perturbation calculation  predicts a nearly scale-invariant spectrum with highly Gaussian distribution (see, e.g., \ \cite{Kodama:1985bj,Mukhanov:1990me} for a review), in good agreement with observations \cite{Akrami:2018vks,Akrami:2018odb}.  In these canonical models, deviations from the scale-invariant Gaussian distribution are controlled by the slow-roll parameters \cite{Maldacena:2002vr}.  Observationally, exact scale-invariant power spectrum has already been ruled out with more than two-$\sigma$ confidence level with the red-tilted spectral index $n_s<1$ \cite{Akrami:2018odb}, but the primordial non-Gaussianity characterized by the bispectrum has not been detected so far, and only constraints on the nonlinearity parameter  $f_{\mathrm{NL}}$ of various  types have been obtained so far \cite{Akrami:2019izv}.

In this situation, a number of extended inflation models have been proposed so far that can realize sizable non-Gaussianity while reproducing the observed red-tilted spectrum.
Theoretically, the simplest local-type non-Gaussianity \cite{Komatsu:2001rj} may be produced by curvaton \cite{PhysRevD.67.023503,PhysRevD.74.103003} or modulated reheating \cite{PhysRevD.69.043508,PhysRevD.77.023505} scenarios, both of which require another fluctuating field in addition to the inflaton. 
Since there is no observational evidence requiring multiple fluctuating fields during inflation \cite{Akrami:2018vks,Akrami:2018odb}, we do not consider such models but stick to the single-field inflation models, whose non-Gaussian signature is mostly the
equilateral one. 
To realize sizable non-Gaussianity in single-field models, some models modify the kinetic terms as in $k$ or $G$ inflation \cite{ArmendarizPicon:1999rj,Garriga:1999vw,Kobayashi:2010cm}, ghost condensate \cite{ArkaniHamed:2003uy,ArkaniHamed:2003uz}, and Dirac-Born Infeld inflation \cite{Alishahiha:2004eh}, while other models extend the gravitational sector \cite{Starobinsky:1979ty,Mijic:1986iv,DeFelice:2010aj,Bamba:2015uma,Baumann:2015xxa}, or both \cite{Kobayashi:2011nu}.  In these noncanonical models, enhanced non-Gaussianity may be realized due to the smallness of sound speed during inflation \cite{Chen:2006nt, Chen:2009bc, Chen:2013aj, Seery:2005wm, Kobayashi:2011pc} and still consistent with the observation \cite{Akrami:2019izv}. Hence primordial non-Gaussianity serves as a good probe of new physics on which these extended models are based \cite{Baumann:2011nk, Baumann:2011su}.

In the language of quantum field theory (QFT), power spectrum of perturbation corresponds to the vacuum expectation value (VEV) of the fluctuation two-point functions. The lowest order of non-Gaussianity is the VEV of the fluctuation three-point functions. Technically, the non-Gaussianity exists because of higher-order interaction terms and it is evaluated by in-in perturbation theory with two-point function as the input to the calculation. However, such interaction terms also generate higher-order corrections to the two-point function which are called loop corrections in QFT terminology.  Such correction must be analyzed carefully to ensure its smallness compared with the tree-level amplitude, or the result of calculations based on the linear perturbation theory, on which most cosmologists rely to set the initial condition of the post-inflationary Universe, would lose its significance.

In this Letter, we calculate one-loop corrections to the power spectrum arising from the three-body interactions of perturbations which also generate primordial non-Gaussianity.  As a result we find that in order to ensure that such loop corrections are small enough to warrant the validity of the conventional lowest-order calculation, the primordial non-Gaussianity must be much smaller than the current observational bound for generic single-field inflation models.

Loop corrections of cosmological perturbations have been of  interest since it was introduced in \cite{Weinberg:2005vy, Weinberg:2006ac}. A number of authors have calculated them for various inflation models with the discussions on  regularization and renormalization schemes \cite{Sloth:2006az, Sloth:2006nu, Seery:2007we, Seery:2007wf, Byrnes:2007tm, Dimastrogiovanni:2008af, Senatore:2009cf, Bartolo:2010bu, Giddings:2010nc, Leblond:2008gg, Assassi:2012et, Melville:2021lst}. Infrared regularity and its relation to quantum states has been discussed in  \cite{Tanaka:2013caa, Urakawa:2010kr}, while loop corrections due to external particles have been discussed in \cite{Chen:2016nrs, Chen:2016uwp}. Loop corrections related to a specific vacuum phase transition model was discussed in \cite{Wu:2017lnh}.

We discuss a single-field inflation model whose Lagrangian is a general function of $X \equiv - \nabla_\mu \phi \nabla^\mu \phi/2$ and $\phi$, namely $P(\phi,X)$, in Einstein gravity. 
It is the simplest inflation model that can generate a large non-Gaussianity. The action of the inflaton is given by 
\begin{equation}
S = \frac{1}{2} \int \mathrm{d}^4x \sqrt{-g} \left[ M_{\mathrm{pl}}^2 R + 2P(X,\phi) \right],
\end{equation}
where $g = \mathrm{det} (g_{\mu \nu})$, $g_{\mu \nu}$ is a metric tensor, and $R$ and $M_{\mathrm{pl}}$ are the Ricci scalar and the reduced Planck scale, respectively. Small perturbation from the homogeneous part of the inflaton $\bar{\phi}(t)$ and metric can be expressed as 
\begin{gather}
\phi(\mathbf{x},t) = \bar{\phi}(t) + \delta \phi(\mathbf{x},t) \nonumber\\
ds^2 = -N^2 \mathrm{d}t^2 + \gamma_{ij} (\mathrm{d}x^i + N^i \mathrm{d}t)(\mathrm{d}x^j + N^j \mathrm{d}t)
\end{gather}
where $\gamma_{ij}$ is a three-dimensional metric on slices of constant $t$, $N$ is a lapse function, and $N^i$ is a shift vector. We work in the comoving gauge
\begin{equation}
\delta \phi(\mathbf{x},t) = 0, ~\gamma_{ij}(\mathbf{x},t) = a^2(t)[1 + 2\zeta(\mathbf{x},t)] \delta_{ij},
\end{equation}
where $N$ and $N^i$ are solutions of Hamiltonian and momentum constraint equations. Also, we do not consider tensor perturbation. After some algebra, the second-order action of the curvature perturbation $\zeta(\mathbf{x},t)$ reads
\begin{equation}
S^{(2)} = M_{\mathrm{pl}}^2 \int \mathrm{d}t ~\mathrm{d}^3x ~a^3 \frac{\epsilon}{c_s^2} \left[ \dot{\zeta}^2 - \frac{c_s^2}{a^2} (\partial_i \zeta)^2  \right],
\label{S2}
\end{equation}
where a dot denotes differentiation with respect to time, and $\epsilon\equiv -\dot{H}/{H^2}$ is a  slow-roll parameter of the Hubble parameter $H$,
which has a very weak time dependence during inflation. $c_s$ is the sound speed expressed as
\begin{equation}
    c_s^2=\frac{P_{,X}}{P_{,X}+2XP_{,XX}},
\end{equation}
with a comma representing a partial derivative.

Before calculating the loop correction to the inflationary power spectrum, we briefly review the standard quantization of cosmological perturbation by introducing the Mukhanov-Sasaki (MS) variable \cite{Mukhanov:1981xt,Sasaki:1986hm}
\begin{equation}
v \equiv M_{\mathrm{pl}}z\zeta , ~~~~z \equiv \frac{a}{c_s}\sqrt{2 \epsilon},
\end{equation}
with which the second-order action (\ref{S2}) becomes canonically normalized
\begin{equation}
S^{(2)} = \frac{1}{2} \int \mathrm{d}\tau ~\mathrm{d}^3x \left[ (v')^2 - c_s^2 (\partial_i v)^2 + \frac{z''}{z} v^2 \right],
\end{equation}
where $\tau$ is conformal time and the prime denotes derivative with respect to $\tau$. During inflation we find, to the lowest order in the slow-roll parameters,
\begin{equation}
    aH=-\frac{1}{(1-\epsilon)\tau},~~a(\tau)\propto \tau^{-(1+\epsilon)},~
    {\rm and}~~\frac{z''}{z}=\frac{2+3\epsilon}{\tau^2}.
\end{equation}
Hence, the operator expansion of the MS variable reads
\begin{equation}
\hat{v}(\mathbf{p}, \tau) =M_{\mathrm{pl}}z\zeta(\mathbf{p}, \tau)=  v_p(\tau) \hat{a}_{\mathbf{p}} + v^*_p (\tau) \hat{a}_{-\mathbf{p}}^\dagger, \nonumber
\end{equation}
with the mode function
\begin{equation}
v_p(\tau) = \left(-\frac{\pi\tau}{4}\right)^{1/2}H_\nu^{(1)}(-p c_s\tau),~~
\nu=\frac{3}{2}+\epsilon,
\end{equation}
\newline
where $\hat{a}_{-\mathbf{p}}^\dagger$ and $\hat{a}_{\mathbf{p}}$ are creation and annihilation operators with the commutation relation
\begin{equation}
\left[ \hat{a}_{\mathbf{p}}, \hat{a}_{-\mathbf{q}}^\dagger \right] = (2 \pi)^3 \delta^3(\mathbf{p} + \mathbf{q}).
\end{equation}

The mode function corresponds to the Bunch-Davies vacuum $\ket{0}$ at early time, which is defined as the state annihilated by $\hat{a}_{\mathbf{p}}$. 
To simplify notation, let us express the two-point function of curvature perturbation and power spectrum at a late time during inflation, $\tau_0$, as
\begin{gather}
\left\langle \zeta(\mathbf{p}) \zeta(\mathbf{q}) \right\rangle = (2 \pi)^3 \delta^3(\mathbf{p} + \mathbf{q}) \left\langle \! \left\langle \zeta(\mathbf{p}) \zeta(-\mathbf{p}) \right\rangle \! \right\rangle , \\
\Delta^2_s(p) \equiv \frac{p^3}{2 \pi^2} \left\langle \! \left\langle \zeta(\mathbf{p}) \zeta(-\mathbf{p}) \right\rangle \! \right\rangle, 
\end{gather}
and take a limit $\tau_0 \rightarrow 0$, when all the relevant modes are in the superhorizon regime and the observed power spectrum is evaluated. Here
the bracket denotes the VEV,  $\langle \cdots \rangle = \bra{0} \cdots \ket{0}$, and $\Delta^2_s(p)$ is the power spectrum multiplied by the phase space density. 

Since the mode function is well described as
\begin{equation}
\zeta_p(\tau) = \frac{v_p (\tau)}{zM_{\mathrm{pl}}} = \left( \frac{H^2}{4M_{\mathrm{pl}}^2 \epsilon c_s} \right)_H^{\frac{1}{2}} \frac{e^{-ic_s p \tau}}{p^{3/2}} (1 + i c_s p \tau), \label{modefunction}
\end{equation}
 the power spectrum is  given by
\begin{equation}
\label{spower}
\Delta^2_{s(0)}(p) = \left( \frac{H^2}{8 \pi^2 M_{\mathrm{pl}}^2 c_s \epsilon} \right)_H = \Delta^2_{s(0)}(p_*) \left( \frac{p}{p_*} \right)^{n_s - 1},
\end{equation}
to the lowest order in slow-roll parameters.
Here subscripts $(0)$ and $H$ denote tree-level contribution and the quantity evaluated at the sound horizon crossing $c_s p = aH$, respectively.
$p_*$ is an arbitrary pivot momentum. The power spectrum is almost scale invariant with the deviation parametrized by the spectral index $n_s - 1 = O(\epsilon)$. Indeed, there is a weak momentum dependence because of time dependence of the quantities in the parenthesis at the horizon crossing of each mode. 

 Beyond second-order action, we can expand more to third-order action. In the limit of $c_s \ll 1$, the third-order action is \cite{Chen:2009bc,Chen:2013aj}
\begin{equation}
S_{\mathrm{int}} = \int \md t ~\md^3 x \left[ - \frac{2 \lambda}{H^3} a^3 \dot{\zeta}^3 + \frac{\epsilon M_{\mathrm{pl}}^2}{ H c_s^2 } a \dot{\zeta} (\partial_i \zeta)^2 \right],  \label{sint}
\end{equation}
where $\lambda = X^2 P_{,XX} + (2/3) X^3 P_{,XXX}$. It is also predicted by the effective field theory of inflation \cite{Cheung:2007st}, where the two interaction terms are independent so observable quantities generated by one  term will not affect the other. Such third-order action is equivalent to cubic self-interaction term of the curvature perturbation. 

 In this Letter, we focus on calculating one-loop correction generated by the second term in \eqref{sint},  which generates primordial non-Gaussianity of the form
\begin{equation}
f_{\mathrm{NL}} = \frac{\langle \zeta \zeta \zeta \rangle}{\langle \zeta \zeta \rangle^2} \propto \frac{1}{c_s^2},
\end{equation}
resulting in  large non-Gaussianity for smaller sound speed.

\begin{widetext}
 Loop correction can be computed by using standard in-in perturbation theory as
\begin{equation}
\langle \mathcal{O(\tau)} \rangle =  \left\langle \left[ \bar{\mathrm{T}} \exp \left( i \int_{-\infty}^{\tau} \mathrm{d}\tau' H_{\mathrm{int}}(\tau') \right) \right] \mathcal{\hat{O}} (\tau) \left[ \mathrm{T} \exp \left( -i \int_{-\infty}^{\tau} \mathrm{d\tau'} H_{\mathrm{int}}(\tau') \right) \right] \right\rangle,
\end{equation}
where $\mathcal{\hat{O}}(\tau)$ is an operator at a fixed time $\tau$, $\mathrm{T}$ and $\bar{\mathrm{T}}$ denote time and antitime ordering. Although the interactions (\ref{sint}) contain the time derivative of the field, the relation $H_{\mathrm{int}} = - \int \mathrm{d}^3x ~\mathcal{L}_{\mathrm{int}}$ still holds with $\mathcal{L}_{\mathrm{int}}$ defined by the integrand of (\ref{sint}) \cite{Adshead:2008gk, Bartolo:2010bu}. In our case, the operator is $\zeta(\mathbf{p}) \zeta(-\mathbf{p})$ evaluated at $\tau=\tau_0 ~(\rightarrow 0)$. First-order expansion vanishes, yielding an odd-point correlation function. 
Second-order expansion of the perturbation theory is
\begin{gather}
\langle \mathcal{O(\tau)} \rangle = \langle \mathcal{O(\tau)} \rangle_{(0,2)}^\dagger + \langle \mathcal{O(\tau)} \rangle_{(1,1)} + \langle \mathcal{O(\tau)} \rangle_{(0,2)}, \nonumber\\
\langle \mathcal{O(\tau)} \rangle_{(1,1)} = \int_{-\infty}^{\tau} \mathrm{d}\tau_1 \int_{-\infty}^{\tau} \mathrm{d}\tau_2 \left\langle H_{\mathrm{int}}(\tau_1) \mathcal{\hat{O}} (\tau) H_{\mathrm{int}}(\tau_2) \right\rangle, \\
\langle \mathcal{O(\tau)} \rangle_{(0,2)} = - \int_{-\infty}^{\tau} \mathrm{d}\tau_1 \int_{-\infty}^{\tau_1} \mathrm{d}\tau_2 \left\langle \mathcal{\hat{O}} (\tau) H_{\mathrm{int}}(\tau_1) H_{\mathrm{int}}(\tau_2) \right\rangle.
\end{gather}
We start the calculation for the $(1,1)$ term by substituting the operator expansion 
\begin{align}
\left\langle \zeta(\bp) \zeta(-\bp) \right\rangle_{(1,1)} = & ~M_{\mathrm{pl}}^4  \int_{-\infty}^0 \md \tau_1 \frac{a^2(\tau_1) \epsilon(\tau_1)}{H(\tau_1) c_s^2} \int_{-\infty}^0 \md \tau_2 \frac{a^2(\tau_2) \epsilon(\tau_2)}{H(\tau_2) c_s^2} \int \prod_{a = 1}^6 \left[ \frac{\md^3 k_a}{(2\pi)^3} \right] \delta^3(\bk_1+\bk_2+\bk_3) ~ \delta^3(\bk_4+\bk_5+\bk_6) \nonumber\\
& (\bk_2 \cdot \bk_3) (\bk_5 \cdot \bk_6) \left\langle \dot{\zeta}(\bk_1,\tau_1) \zeta(\bk_2,\tau_1) \zeta(\bk_3,\tau_1) \zeta(\bp) \zeta(-\bp) \dot{\zeta}(\bk_4,\tau_2) \zeta(\bk_5,\tau_2) \zeta(\bk_6,\tau_2) \right\rangle.
\end{align}
For small loop momentum $k \ll p$, performing Wick contraction and substituting mode function yields
\begin{align}
\left\langle \! \left\langle \zeta(\mathbf{p}) \zeta(-\mathbf{p}) \right\rangle \! \right\rangle_{(1,1)} = 2p^4 \left\langle \! \left\langle \zeta(\mathbf{p}) \zeta(-\mathbf{p}) \right\rangle \! \right\rangle_{(0)} & \int_{-\infty}^0 \md \tau_1 \int_{-\infty}^0 \md \tau_2 (1 + i c_s p \tau_1)^2 (1 - i c_s p \tau_2)^2 e^{-2i c_s p(\tau_1-\tau_2)} \nonumber\\
& \int \frac{\md^3 k}{(2\pi)^3} \frac{k}{4c_s^2 p^6} \left( \frac{H^2}{4 M_{\mathrm{pl}}^2 \epsilon c_s} \right)_H.
\end{align}
We find that contribution from small loop momentum in the limit $k \rightarrow 0$ is zero, consistent with Maldacena's theorem \cite{Maldacena:2002vr}. In the same limit, the $(0,2)$ term will generate the same momentum dependence.

Hence, the loop correction comes from other domain, namely, $k > p$. In this domain, the divergent integral comes from the $(0,2)$ term, while the $(1,1)$ term converges in a similar way to \cite{Senatore:2009cf}. The $(0,2)$ term is given by
\begin{align}
\left\langle \zeta(\bp) \zeta(-\bp) \right\rangle_{(0,2)} = & -M_{\mathrm{pl}}^4 \int_{-\infty}^0 \md \tau_1 \frac{a^2(\tau_1) \epsilon(\tau_1)}{H(\tau_1) c_s^2} \int_{-\infty}^{\tau_1} \md \tau_2 \frac{a^2(\tau_2) \epsilon(\tau_2)}{H(\tau_2) c_s^2} \int \prod_{a = 1}^6 \left[ \frac{\md^3 k_a}{(2\pi)^3} \right] \delta^3(\bk_1+\bk_2+\bk_3) ~ \delta^3(\bk_4+\bk_5+\bk_6) \nonumber\\
& (\bk_2 \cdot \bk_3) (\bk_5 \cdot \bk_6) \left\langle \zeta(\bp) \zeta(-\bp) \dot{\zeta}(\bk_1,\tau_1) \zeta(\bk_2,\tau_1) \zeta(\bk_3,\tau_1) \dot{\zeta}(\bk_4,\tau_2) \zeta(\bk_5,\tau_2) \zeta(\bk_6,\tau_2) \right\rangle.  \label{start}
\end{align}
Performing Wick contraction, it becomes
\begin{equation}
\left\langle \! \left\langle \zeta(\mathbf{p}) \zeta(-\mathbf{p}) \right\rangle \! \right\rangle_{(0,2)} = -M_{\mathrm{pl}}^4  \int_{-\infty}^0 \md \tau_1 \frac{a^2(\tau_1) \epsilon(\tau_1)}{H(\tau_1) c_s^2} \int_{-\infty}^{\tau_1} \md \tau_2 \frac{a^2(\tau_2) \epsilon(\tau_2)}{H(\tau_2) c_s^2} \zeta_p(\tau_0) \zeta_p(\tau_0) f(p, \tau_1, \tau_2), \label{pmp}
\end{equation}
\begin{align}
f(p, \tau_1, \tau_2) &= \int \frac{\md^3k}{(2\pi)^3}  \left[ 8(\bp \cdot \bq)^2  \dot{\zeta}_k(\tau_1) \dot{\zeta}^*_k(\tau_2) \zeta_p^*(\tau_1) \zeta_p^*(\tau_2) \zeta_q(\tau_1) \zeta_q^*(\tau_2)  \right. \label{ef} \\
& \left. -8 (\bp \cdot \bk) (\bp \cdot \bq) \dot{\zeta}_k(\tau_1) \dot{\zeta}^*_q(\tau_2) \zeta_p^*(\tau_1) \zeta_p^*(\tau_2) \zeta_q(\tau_1) \zeta_k^*(\tau_2)  -8 (\bp \cdot \bq) (\bk \cdot \bq) \dot{\zeta}_k(\tau_1) \dot{\zeta}^*_p(\tau_2) \zeta_p^*(\tau_1) \zeta_k^*(\tau_2) \zeta_q(\tau_1) \zeta_q^*(\tau_2)  \right. \nonumber\\
& \left. -8 (\bp \cdot \bq) (\bk \cdot \bq) \dot{\zeta}^*_p(\tau_1) \dot{\zeta}^*_k(\tau_2) \zeta_k(\tau_1) \zeta_p^*(\tau_2) \zeta_q(\tau_1) \zeta_q^*(\tau_2)  + 4(\bk \cdot \bq)^2  \dot{\zeta}^*_p(\tau_1) \dot{\zeta}^*_p(\tau_2) \zeta_k(\tau_1) \zeta_k^*(\tau_2) \zeta_q(\tau_1) \zeta_q^*(\tau_2) \right], \nonumber
\end{align}
where $\bq = \bk -\bp$. Then, substituting mode function and performing time integration, 
we find ultraviolet (UV) divergent terms with different power of $k$. The terms that diverge with a positive power of $k$ are to be regularized and renormalized in the same way as QFT in flat spacetime, and are irrelevant in cosmological consideration.
The most important term here is the term with its $k$-integrand proportional to $k^{-3}$, which would yield a logarithmic divergence in pure dS background with a scale-invariant spectrum \cite{Senatore:2009cf, Bartolo:2010bu}. Fortunately, however, in a realistic situation with
a red-tilted power spectrum, such a term is multiplied by a function with a weak $k$ dependence suppressed by a slow-roll parameter, which automatically removes the logarithmic divergence as we see now.

Extracting terms with the above mentioned $k$ dependence from
(\ref{pmp}) yields
\begin{align}
\left\langle \! \left\langle \zeta(\mathbf{p}) \zeta(-\mathbf{p}) \right\rangle \! \right\rangle_{(0,2)} &\cong \frac{51}{80 c_s^4} \left\langle \! \left\langle \zeta(\mathbf{p}) \zeta(-\mathbf{p}) \right\rangle \! \right\rangle_{(0)} \int_p^\infty \md k ~k^2 \frac{1}{k^3} \left( \frac{H^2}{8 \pi^2 M_{\mathrm{pl}}^2 \epsilon c_s} \right)_H \nonumber\\
&= \frac{51}{80 c_s^4} \left\langle \! \left\langle \zeta(\mathbf{p}) \zeta(-\mathbf{p}) \right\rangle \! \right\rangle_{(0)} \int_p^\infty \md k ~k^2 \frac{1}{k^3} \Delta^2_{s(0)}(k_*) \left( \frac{k}{k_*} \right)^{n_s-1}. \label{loop}
\end{align}
For $n_s < 1$, which is the case in our Universe, the momentum integration converges as
\begin{equation}
\left\langle \! \left\langle \zeta(\mathbf{p}) \zeta(-\mathbf{p}) \right\rangle \! \right\rangle_{(0,2)} = \frac{51}{80 c_s^4} \left\langle \! \left\langle \zeta(\mathbf{p}) \zeta(-\mathbf{p}) \right\rangle \! \right\rangle_{(0)} \frac{\Delta^2_{s(0)}(p)}{1 - n_s}.
\end{equation}
\end{widetext}
Hence, adding the contribution of the $(2,0)$ term, the total one-loop correction to the power spectrum is
\begin{equation}
\Delta^2_{s(1)}(p) = \frac{ 51 [ \Delta^2_{s(0)}(p) ]^2}{40 c_s^4 (1 - n_s)}.  \label{correction}
\end{equation}

Note that preceding analyses \cite{Senatore:2009cf, Bartolo:2010bu}, which studied the same type of loop correction, did not take the spectral tilt into account, but stuck to the scale-invariant spectrum. As a consequence, they encountered a logarithmic divergence and introduced dimensional regularization
\begin{equation}
\int \frac{\md^3 k}{(2\pi)^3 k^3} \rightarrow \int \frac{\md^{3+\delta} k}{(2\pi)^3 k^3},
\end{equation}
where $\delta$ is a very small arbitrary parameter. Then, they got a $1/\delta$ pole and renormalize it to make the observable free of an arbitrary parameter. In our observed Universe, however, the spectrum has a red tilt, which makes the momentum integral  convergent. Therefore, we do not need regularization and renormalization introducing an extra parameter $\delta$, which should not be confused with a physical observable $1- n_s$. 

In order for the perturbative calculation to be credible, one-loop correction must be suppressed compared with the tree-level contribution, otherwise the power spectrum would become infinitely large if we calculate higher-order loop correction. Thus we require
\begin{equation}
\frac{\Delta_{s(0)}^2}{c_s^4 (1-n_s)} \ll 1.   \label{33}
\end{equation}
We now estimate the left-hand side of the above inequality from observational results \cite{Akrami:2018odb}.  Although $\Delta_{s(0)}^2$ corresponds to the "bare" quantity in QFT language, let us put the observed amplitude of the scalar power spectrum at the pivot scale $k_\ast=0.05~\mathrm{Mpc}^{-1}$, $\Delta_{s(0)}^2\equiv A_2=2.1\times 10^{-9}$ \cite{Akrami:2018odb} according to the common practice of cosmology which we are about to refine. From the observed value of the spectral index, $n_s=0.9649 \pm 0.0042$ based on $TT$, $TE$, $EE$, low $E$, and lens of Planck 2018 \cite{Akrami:2018odb}, let us take a conservative value $1-n_s=0.0393$ from its  one-$\sigma$ lower bound. Then the inequality (\ref{33}) imposes a constraint on the sound speed as $c_s \gg 0.02$. 

Interestingly, this inequality practically coincides with the observational bound on the sound speed imposed by the equilateral non-Gaussianity \cite{Akrami:2019izv}, namely, $c_s> 0.021$ at 95\% confidence level. Since our inequality is a strong one with $\gg$, this means that the amplitude of the actual non-Gaussianity must be much smaller than the current observational bound; otherwise the conventional cosmological perturbation theory, which has been successfully describing the evolution of the Universe in agreement with observations, would lose the predictability \footnote{Here one may recall Einstein's saying that the eternal mystery of the world is its comprehensibility.}, running into the strong coupling regime \cite{Baumann:2011dt}. 

If the lower limit of sound speed is increased to $c_s \sim 0.1$, it will be safe from a dangerous loop correction as it is suppressed to $O(0.001)$, but it is still as large as slow-roll parameter corrections to the power spectrum \cite{Stewart:1993bc}. 

We may compare our result to the constraint presented in \cite{Bartolo:2010bu}, where they considered an exact scale-invariant power spectrum so they did not get observationally relevant inverse of the $1-n_s$ factor. A similar constraint was also discussed in \cite{Leblond:2008gg}, where they only analyzed the coupling constant of the interaction action. As we have shown here, however, proper incorporation of background dynamics is essentially important to get the enhancement factor. Therefore, our constraint is more severe and places primordial non-Gaussianity in a stricter limit. This is important as non-Gaussianity has been proposed as the probe of new fundamental physics.

So far we have assumed a single power-law spectrum in the entire wave number range.  If we relax this assumption and consider a wider possibility of the shape of the spectrum on smaller scales that cannot be probed by CMB observations, we may find various interesting implications to cosmology of the early Universe.  For example, if small-scale perturbations with wave number $p'$ have a different amplitude and red spectral index $n'_s$, then the condition \eqref{33} may rather be interpreted as a constraint on the amplitude of fluctuations
\begin{equation}
\Delta_{s(0)}^2(p') \ll c_s^4(1 - n'_s) \approx f_{\mathrm{NL}}^{-2} (1 - n'_s).
\end{equation}
For $c_s= 0.1$ and $1-n'_s=0.1$, for example, we find $\Delta_{s(0)}^2(p') \ll 10^{-5}$, which is already
more stringent than some known constraints 
\cite{Jeong:2014gna,Nakama:2014vla}.

More interesting is the case in which the power spectrum has some peaky shape, which may induce formation of primordial black holes (PBHs) \cite{1967SvA....10..602Z,Hawking:1974rv,Carr:2009jm,Carr:2020gox} and large stochastic gravitational wave background \cite{Baumann:2007zm,Assadullahi:2009jc,Saito:2008jc,Saito:2009jt}. Since the shape of the power spectrum as well as its generation mechanism vary model to model, let us consider a simple toy model where $\zeta_p(t)$ has the same functional form as \eqref{modefunction}, but is enhanced by a constant factor $B \gg 1$ for a finite wave number range from $k_p$ to $k_p e^{N_p}$ where $N_p={\cal O}(1)$.  Repeating the same calculations from \eqref{start} to \eqref{loop}, we find that the wave number integral in \eqref{ef} is multiplied by a factor $B^4$ for $k_p<k<k_pe^{N_p}$,
assuming $k_p \gg p$, so that \eqref{correction}
acquires an extra component,
\begin{equation}
\Delta^2_{s(1)}(p) = \frac{ 51 [ \Delta^2_{s(0)}(p) ]^2}{40 c_s^4 (1 - n_s)}
+ {\cal O}\left( \frac{N_p}{c_s^4}[ \Delta^2_{s(0)}(k_p) ]^2\right).  \label{peaky}
\end{equation}
Then the perturbativity condition reads
\begin{equation}
  \Delta^2_{s(0)}(k_p) \ll \left(\frac{c_s^4}{N_p}\Delta^2_{s(0)}(p) \right)^{\frac{1}{2}}
  \sim (5\times 10^{-5})c_s^2.
\end{equation}
If we take this constraint at face value, formation of an appreciable amount of PBHs nor observable amplitude of stochastic gravitational wave background are by no means possible unless deci-hertz interferometer gravitational wave observatory (DECIGO) or big bang observer (BBO) are realized for the latter \cite{Assadullahi:2009jc}. 

In fact, however, this result should be interpreted with caveats, as it has been derived under a number of assumptions such as the form of interaction \eqref{sint} taking $c_s$ small and an significantly simplified mode function. Nevertheless, this simple analysis certainly motivates us to study the constraint in more realistic single-field inflation models  predicting PBH formation (e.g., \cite{Ivanov:1994pa,Saito:2008em,Motohashi:2017kbs}) or models inducing large-amplitude stochastic gravitational wave background \cite{Cai:2018dig}
to which our analysis does not directly apply because these authors assumed a different type of non-Gaussianity than we consider here.

Consistency with Maldacena's theorem implies that contribution from small loop momentum to the one-loop correction \eqref{loop} is zero, 
which we have confirmed by direct manipulations. However, there are some known single-field examples that violate Maldacena's consistency condition, such as the ultra slow-roll (USR) inflation \cite{Kinney:2005vj,Martin:2012pe,Motohashi:2014ppa}.  One may wonder if such models suffer from infrared divergence, as they are not protected by Maldacena's theorem.  Since such models are realized as a nonattractor solution, which does not last long,  modes relevant to the USR regime span only a finite range of wave number \cite{Suyama:2021adn}, and hence they are free from the infrared problem.
These modes, however, may contribute significantly to the one-loop corrections, as
the would-be decaying mode of perturbation may grow during such regime in a similar way to \cite{Saito:2008em}.  In such a case, the mode function has a different form from \eqref{modefunction}, so a separate analysis is required. Note that the non-Gaussianity generated in this model has a different shape, too \cite{Firouzjahi:2018vet,Namjoo:2012aa}.

In conclusion, we calculated the one-loop correction to the inflationary power spectrum by using standard in-in perturbation theory. The red tilt of the spectral index makes the loop's momentum integration convergent. Because of it, the loop correction is enhanced by an inverse factor of $1-n_s$, which is a small positive number according to the latest observation \cite{Akrami:2018odb}.
As a result, in order for the loop correction to be small enough to warrant the validity of the standard perturbation theory, the amplitude of equilateral non-Gaussianity must be much smaller than the current observational bound for the case where the primordial spectrum has a simple power-law spectrum. This consideration opens up the possibility of further interesting consequences for a more general spectrum, which will be reported elsewhere.

We thank Enrico Pajer for a very helpful discussion about Maldacena's consistency condition. We also thank Yusuke Yamada and Kohei Kamada for the discussion in the early stage of writing this Letter. J. K. acknowledges the support from Monbukagakusho (MEXT) scholarship and Global Science Graduate Course (GSGC) program of The University of Tokyo. J. Y. is supported by JSPS KAKENHI Grant No.~20H05639 and Grant-in-Aid for Scientific Research on Innovative Areas 20H05248. 

\bibliographystyle{apsrev4-1}
\bibliography{Reference}

\providecommand{\noopsort}[1]{}\providecommand{\singleletter}[1]{#1}%
\begin{thebibliography}{83}%
\makeatletter
\providecommand \@ifxundefined [1]{%
 \@ifx{#1\undefined}
}%
\providecommand \@ifnum [1]{%
 \ifnum #1\expandafter \@firstoftwo
 \else \expandafter \@secondoftwo
 \fi
}%
\providecommand \@ifx [1]{%
 \ifx #1\expandafter \@firstoftwo
 \else \expandafter \@secondoftwo
 \fi
}%
\providecommand \natexlab [1]{#1}%
\providecommand \enquote  [1]{``#1''}%
\providecommand \bibnamefont  [1]{#1}%
\providecommand \bibfnamefont [1]{#1}%
\providecommand \citenamefont [1]{#1}%
\providecommand \href@noop [0]{\@secondoftwo}%
\providecommand \href [0]{\begingroup \@sanitize@url \@href}%
\providecommand \@href[1]{\@@startlink{#1}\@@href}%
\providecommand \@@href[1]{\endgroup#1\@@endlink}%
\providecommand \@sanitize@url [0]{\catcode `\\12\catcode `\$12\catcode
  `\&12\catcode `\#12\catcode `\^12\catcode `\_12\catcode `\%12\relax}%
\providecommand \@@startlink[1]{}%
\providecommand \@@endlink[0]{}%
\providecommand \url  [0]{\begingroup\@sanitize@url \@url }%
\providecommand \@url [1]{\endgroup\@href {#1}{\urlprefix }}%
\providecommand \urlprefix  [0]{URL }%
\providecommand \Eprint [0]{\href }%
\providecommand \doibase [0]{http://dx.doi.org/}%
\providecommand \selectlanguage [0]{\@gobble}%
\providecommand \bibinfo  [0]{\@secondoftwo}%
\providecommand \bibfield  [0]{\@secondoftwo}%
\providecommand \translation [1]{[#1]}%
\providecommand \BibitemOpen [0]{}%
\providecommand \bibitemStop [0]{}%
\providecommand \bibitemNoStop [0]{.\EOS\space}%
\providecommand \EOS [0]{\spacefactor3000\relax}%
\providecommand \BibitemShut  [1]{\csname bibitem#1\endcsname}%
\let\auto@bib@innerbib\@empty
%</preamble>
\bibitem [{\citenamefont {Starobinsky}(1980)}]{Starobinsky:1980te}%
  \BibitemOpen
  \bibfield  {author} {\bibinfo {author} {\bibfnamefont {A.~A.}\ \bibnamefont
  {Starobinsky}},\ }\href {\doibase 10.1016/0370-2693(80)90670-X} {\bibfield
  {journal} {\bibinfo  {journal} {Phys. Lett. B}\ }\textbf {\bibinfo {volume}
  {91}},\ \bibinfo {pages} {99} (\bibinfo {year} {1980})}\BibitemShut {NoStop}%
\bibitem [{\citenamefont {Sato}(1981)}]{Sato:1980yn}%
  \BibitemOpen
  \bibfield  {author} {\bibinfo {author} {\bibfnamefont {K.}~\bibnamefont
  {Sato}},\ }\href@noop {} {\bibfield  {journal} {\bibinfo  {journal} {Mon.
  Not. Roy. Astron. Soc.}\ }\textbf {\bibinfo {volume} {195}},\ \bibinfo
  {pages} {467} (\bibinfo {year} {1981})}\BibitemShut {NoStop}%
\bibitem [{\citenamefont {Guth}(1980)}]{Guth:1980zm}%
  \BibitemOpen
  \bibfield  {author} {\bibinfo {author} {\bibfnamefont {A.~H.}\ \bibnamefont
  {Guth}},\ }\href {\doibase 10.1103/PhysRevD.23.347} {\bibfield  {journal}
  {\bibinfo  {journal} {Phys. Rev.}\ }\textbf {\bibinfo {volume} {D23}},\
  \bibinfo {pages} {347} (\bibinfo {year} {1980})}\BibitemShut {NoStop}%
\bibitem [{\citenamefont {Sato}\ and\ \citenamefont
  {Yokoyama}(2015)}]{Sato:2015dga}%
  \BibitemOpen
  \bibfield  {author} {\bibinfo {author} {\bibfnamefont {K.}~\bibnamefont
  {Sato}}\ and\ \bibinfo {author} {\bibfnamefont {J.}~\bibnamefont
  {Yokoyama}},\ }\href {\doibase 10.1142/S0218271815300256} {\bibfield
  {journal} {\bibinfo  {journal} {Int. J. Mod. Phys. D}\ }\textbf {\bibinfo
  {volume} {24}},\ \bibinfo {pages} {1530025} (\bibinfo {year}
  {2015})}\BibitemShut {NoStop}%
\bibitem [{\citenamefont {Starobinsky}(1979)}]{Starobinsky:1979ty}%
  \BibitemOpen
  \bibfield  {author} {\bibinfo {author} {\bibfnamefont {A.~A.}\ \bibnamefont
  {Starobinsky}},\ }\href@noop {} {\bibfield  {journal} {\bibinfo  {journal}
  {JETP Lett.}\ }\textbf {\bibinfo {volume} {30}},\ \bibinfo {pages} {682}
  (\bibinfo {year} {1979})}\BibitemShut {NoStop}%
\bibitem [{\citenamefont {Mukhanov}\ and\ \citenamefont
  {Chibisov}(1981)}]{Mukhanov:1981xt}%
  \BibitemOpen
  \bibfield  {author} {\bibinfo {author} {\bibfnamefont {V.~F.}\ \bibnamefont
  {Mukhanov}}\ and\ \bibinfo {author} {\bibfnamefont {G.~V.}\ \bibnamefont
  {Chibisov}},\ }\href@noop {} {\bibfield  {journal} {\bibinfo  {journal} {JETP
  Lett.}\ }\textbf {\bibinfo {volume} {33}},\ \bibinfo {pages} {532} (\bibinfo
  {year} {1981})}\BibitemShut {NoStop}%
\bibitem [{\citenamefont {Starobinsky}(1982)}]{Starobinsky:1982ee}%
  \BibitemOpen
  \bibfield  {author} {\bibinfo {author} {\bibfnamefont {A.~A.}\ \bibnamefont
  {Starobinsky}},\ }\href {\doibase 10.1016/0370-2693(82)90541-X} {\bibfield
  {journal} {\bibinfo  {journal} {Phys. Lett. B}\ }\textbf {\bibinfo {volume}
  {117}},\ \bibinfo {pages} {175} (\bibinfo {year} {1982})}\BibitemShut
  {NoStop}%
\bibitem [{\citenamefont {Hawking}(1982)}]{Hawking:1982cz}%
  \BibitemOpen
  \bibfield  {author} {\bibinfo {author} {\bibfnamefont {S.~W.}\ \bibnamefont
  {Hawking}},\ }\href {\doibase 10.1016/0370-2693(82)90373-2} {\bibfield
  {journal} {\bibinfo  {journal} {Phys. Lett. B}\ }\textbf {\bibinfo {volume}
  {115}},\ \bibinfo {pages} {295} (\bibinfo {year} {1982})}\BibitemShut
  {NoStop}%
\bibitem [{\citenamefont {Guth}\ and\ \citenamefont {Pi}(1982)}]{Guth:1982ec}%
  \BibitemOpen
  \bibfield  {author} {\bibinfo {author} {\bibfnamefont {A.~H.}\ \bibnamefont
  {Guth}}\ and\ \bibinfo {author} {\bibfnamefont {S.~Y.}\ \bibnamefont {Pi}},\
  }\href {\doibase 10.1103/PhysRevLett.49.1110} {\bibfield  {journal} {\bibinfo
   {journal} {Phys. Rev. Lett.}\ }\textbf {\bibinfo {volume} {49}},\ \bibinfo
  {pages} {1110} (\bibinfo {year} {1982})}\BibitemShut {NoStop}%
\bibitem [{\citenamefont {Kodama}\ and\ \citenamefont
  {Sasaki}(1984)}]{Kodama:1985bj}%
  \BibitemOpen
  \bibfield  {author} {\bibinfo {author} {\bibfnamefont {H.}~\bibnamefont
  {Kodama}}\ and\ \bibinfo {author} {\bibfnamefont {M.}~\bibnamefont
  {Sasaki}},\ }\href {\doibase 10.1143/PTPS.78.1} {\bibfield  {journal}
  {\bibinfo  {journal} {Prog. Theor. Phys. Suppl.}\ }\textbf {\bibinfo {volume}
  {78}},\ \bibinfo {pages} {1} (\bibinfo {year} {1984})}\BibitemShut {NoStop}%
\bibitem [{\citenamefont {Mukhanov}\ \emph {et~al.}(1992)\citenamefont
  {Mukhanov}, \citenamefont {Feldman},\ and\ \citenamefont
  {Brandenberger}}]{Mukhanov:1990me}%
  \BibitemOpen
  \bibfield  {author} {\bibinfo {author} {\bibfnamefont {V.~F.}\ \bibnamefont
  {Mukhanov}}, \bibinfo {author} {\bibfnamefont {H.~A.}\ \bibnamefont
  {Feldman}}, \ and\ \bibinfo {author} {\bibfnamefont {R.~H.}\ \bibnamefont
  {Brandenberger}},\ }\href {\doibase 10.1016/0370-1573(92)90044-Z} {\bibfield
  {journal} {\bibinfo  {journal} {Phys. Rept.}\ }\textbf {\bibinfo {volume}
  {215}},\ \bibinfo {pages} {203} (\bibinfo {year} {1992})}\BibitemShut
  {NoStop}%
\bibitem [{\citenamefont {Aghanim}\ \emph {et~al.}(2020)\citenamefont {Aghanim}
  \emph {et~al.}}]{Akrami:2018vks}%
  \BibitemOpen
  \bibfield  {author} {\bibinfo {author} {\bibfnamefont {N.}~\bibnamefont
  {Aghanim}} \emph {et~al.} (\bibinfo {collaboration} {Planck}),\ }\href
  {\doibase 10.1051/0004-6361/201833880} {\bibfield  {journal} {\bibinfo
  {journal} {Astron. Astrophys.}\ }\textbf {\bibinfo {volume} {641}},\ \bibinfo
  {pages} {A1} (\bibinfo {year} {2020})},\ \Eprint
  {http://arxiv.org/abs/1807.06205} {arXiv:1807.06205 [astro-ph.CO]}
  \BibitemShut {NoStop}%
\bibitem [{\citenamefont {Akrami}\ \emph
  {et~al.}(2020{\natexlab{a}})\citenamefont {Akrami} \emph
  {et~al.}}]{Akrami:2018odb}%
  \BibitemOpen
  \bibfield  {author} {\bibinfo {author} {\bibfnamefont {Y.}~\bibnamefont
  {Akrami}} \emph {et~al.} (\bibinfo {collaboration} {Planck}),\ }\href
  {\doibase 10.1051/0004-6361/201833887} {\bibfield  {journal} {\bibinfo
  {journal} {Astron. Astrophys.}\ }\textbf {\bibinfo {volume} {641}},\ \bibinfo
  {pages} {A10} (\bibinfo {year} {2020}{\natexlab{a}})},\ \Eprint
  {http://arxiv.org/abs/1807.06211} {arXiv:1807.06211 [astro-ph.CO]}
  \BibitemShut {NoStop}%
\bibitem [{\citenamefont {Maldacena}(2003)}]{Maldacena:2002vr}%
  \BibitemOpen
  \bibfield  {author} {\bibinfo {author} {\bibfnamefont {J.~M.}\ \bibnamefont
  {Maldacena}},\ }\href {\doibase 10.1088/1126-6708/2003/05/013} {\bibfield
  {journal} {\bibinfo  {journal} {JHEP}\ }\textbf {\bibinfo {volume} {05}},\
  \bibinfo {pages} {013} (\bibinfo {year} {2003})},\ \Eprint
  {http://arxiv.org/abs/astro-ph/0210603} {arXiv:astro-ph/0210603} \BibitemShut
  {NoStop}%
\bibitem [{\citenamefont {Akrami}\ \emph
  {et~al.}(2020{\natexlab{b}})\citenamefont {Akrami} \emph
  {et~al.}}]{Akrami:2019izv}%
  \BibitemOpen
  \bibfield  {author} {\bibinfo {author} {\bibfnamefont {Y.}~\bibnamefont
  {Akrami}} \emph {et~al.} (\bibinfo {collaboration} {Planck}),\ }\href
  {\doibase 10.1051/0004-6361/201935891} {\bibfield  {journal} {\bibinfo
  {journal} {Astron. Astrophys.}\ }\textbf {\bibinfo {volume} {641}},\ \bibinfo
  {pages} {A9} (\bibinfo {year} {2020}{\natexlab{b}})},\ \Eprint
  {http://arxiv.org/abs/1905.05697} {arXiv:1905.05697 [astro-ph.CO]}
  \BibitemShut {NoStop}%
\bibitem [{\citenamefont {Komatsu}\ and\ \citenamefont
  {Spergel}(2001)}]{Komatsu:2001rj}%
  \BibitemOpen
  \bibfield  {author} {\bibinfo {author} {\bibfnamefont {E.}~\bibnamefont
  {Komatsu}}\ and\ \bibinfo {author} {\bibfnamefont {D.~N.}\ \bibnamefont
  {Spergel}},\ }\href {\doibase 10.1103/PhysRevD.63.063002} {\bibfield
  {journal} {\bibinfo  {journal} {Phys. Rev. D}\ }\textbf {\bibinfo {volume}
  {63}},\ \bibinfo {pages} {063002} (\bibinfo {year} {2001})},\ \Eprint
  {http://arxiv.org/abs/astro-ph/0005036} {arXiv:astro-ph/0005036} \BibitemShut
  {NoStop}%
\bibitem [{\citenamefont {Lyth}\ \emph {et~al.}(2003)\citenamefont {Lyth},
  \citenamefont {Ungarelli},\ and\ \citenamefont {Wands}}]{PhysRevD.67.023503}%
  \BibitemOpen
  \bibfield  {author} {\bibinfo {author} {\bibfnamefont {D.~H.}\ \bibnamefont
  {Lyth}}, \bibinfo {author} {\bibfnamefont {C.}~\bibnamefont {Ungarelli}}, \
  and\ \bibinfo {author} {\bibfnamefont {D.}~\bibnamefont {Wands}},\ }\href
  {\doibase 10.1103/PhysRevD.67.023503} {\bibfield  {journal} {\bibinfo
  {journal} {Phys. Rev. D}\ }\textbf {\bibinfo {volume} {67}},\ \bibinfo
  {pages} {023503} (\bibinfo {year} {2003})}\BibitemShut {NoStop}%
\bibitem [{\citenamefont {Sasaki}\ \emph {et~al.}(2006)\citenamefont {Sasaki},
  \citenamefont {V\"aliviita},\ and\ \citenamefont
  {Wands}}]{PhysRevD.74.103003}%
  \BibitemOpen
  \bibfield  {author} {\bibinfo {author} {\bibfnamefont {M.}~\bibnamefont
  {Sasaki}}, \bibinfo {author} {\bibfnamefont {J.}~\bibnamefont {V\"aliviita}},
  \ and\ \bibinfo {author} {\bibfnamefont {D.}~\bibnamefont {Wands}},\ }\href
  {\doibase 10.1103/PhysRevD.74.103003} {\bibfield  {journal} {\bibinfo
  {journal} {Phys. Rev. D}\ }\textbf {\bibinfo {volume} {74}},\ \bibinfo
  {pages} {103003} (\bibinfo {year} {2006})}\BibitemShut {NoStop}%
\bibitem [{\citenamefont {Zaldarriaga}(2004)}]{PhysRevD.69.043508}%
  \BibitemOpen
  \bibfield  {author} {\bibinfo {author} {\bibfnamefont {M.}~\bibnamefont
  {Zaldarriaga}},\ }\href {\doibase 10.1103/PhysRevD.69.043508} {\bibfield
  {journal} {\bibinfo  {journal} {Phys. Rev. D}\ }\textbf {\bibinfo {volume}
  {69}},\ \bibinfo {pages} {043508} (\bibinfo {year} {2004})}\BibitemShut
  {NoStop}%
\bibitem [{\citenamefont {Suyama}\ and\ \citenamefont
  {Yamaguchi}(2008)}]{PhysRevD.77.023505}%
  \BibitemOpen
  \bibfield  {author} {\bibinfo {author} {\bibfnamefont {T.}~\bibnamefont
  {Suyama}}\ and\ \bibinfo {author} {\bibfnamefont {M.}~\bibnamefont
  {Yamaguchi}},\ }\href {\doibase 10.1103/PhysRevD.77.023505} {\bibfield
  {journal} {\bibinfo  {journal} {Phys. Rev. D}\ }\textbf {\bibinfo {volume}
  {77}},\ \bibinfo {pages} {023505} (\bibinfo {year} {2008})}\BibitemShut
  {NoStop}%
\bibitem [{\citenamefont {Armendariz-Picon}\ \emph {et~al.}(1999)\citenamefont
  {Armendariz-Picon}, \citenamefont {Damour},\ and\ \citenamefont
  {Mukhanov}}]{ArmendarizPicon:1999rj}%
  \BibitemOpen
  \bibfield  {author} {\bibinfo {author} {\bibfnamefont {C.}~\bibnamefont
  {Armendariz-Picon}}, \bibinfo {author} {\bibfnamefont {T.}~\bibnamefont
  {Damour}}, \ and\ \bibinfo {author} {\bibfnamefont {V.~F.}\ \bibnamefont
  {Mukhanov}},\ }\href {\doibase 10.1016/S0370-2693(99)00603-6} {\bibfield
  {journal} {\bibinfo  {journal} {Phys. Lett. B}\ }\textbf {\bibinfo {volume}
  {458}},\ \bibinfo {pages} {209} (\bibinfo {year} {1999})},\ \Eprint
  {http://arxiv.org/abs/hep-th/9904075} {arXiv:hep-th/9904075} \BibitemShut
  {NoStop}%
\bibitem [{\citenamefont {Garriga}\ and\ \citenamefont
  {Mukhanov}(1999)}]{Garriga:1999vw}%
  \BibitemOpen
  \bibfield  {author} {\bibinfo {author} {\bibfnamefont {J.}~\bibnamefont
  {Garriga}}\ and\ \bibinfo {author} {\bibfnamefont {V.~F.}\ \bibnamefont
  {Mukhanov}},\ }\href {\doibase 10.1016/S0370-2693(99)00602-4} {\bibfield
  {journal} {\bibinfo  {journal} {Phys. Lett. B}\ }\textbf {\bibinfo {volume}
  {458}},\ \bibinfo {pages} {219} (\bibinfo {year} {1999})},\ \Eprint
  {http://arxiv.org/abs/hep-th/9904176} {arXiv:hep-th/9904176} \BibitemShut
  {NoStop}%
\bibitem [{\citenamefont {Kobayashi}\ \emph {et~al.}(2010)\citenamefont
  {Kobayashi}, \citenamefont {Yamaguchi},\ and\ \citenamefont
  {Yokoyama}}]{Kobayashi:2010cm}%
  \BibitemOpen
  \bibfield  {author} {\bibinfo {author} {\bibfnamefont {T.}~\bibnamefont
  {Kobayashi}}, \bibinfo {author} {\bibfnamefont {M.}~\bibnamefont
  {Yamaguchi}}, \ and\ \bibinfo {author} {\bibfnamefont {J.}~\bibnamefont
  {Yokoyama}},\ }\href {\doibase 10.1103/PhysRevLett.105.231302} {\bibfield
  {journal} {\bibinfo  {journal} {Phys. Rev. Lett.}\ }\textbf {\bibinfo
  {volume} {105}},\ \bibinfo {pages} {231302} (\bibinfo {year} {2010})},\
  \Eprint {http://arxiv.org/abs/1008.0603} {arXiv:1008.0603 [hep-th]}
  \BibitemShut {NoStop}%
\bibitem [{\citenamefont {Arkani-Hamed}\ \emph
  {et~al.}(2004{\natexlab{a}})\citenamefont {Arkani-Hamed}, \citenamefont
  {Cheng}, \citenamefont {Luty},\ and\ \citenamefont
  {Mukohyama}}]{ArkaniHamed:2003uy}%
  \BibitemOpen
  \bibfield  {author} {\bibinfo {author} {\bibfnamefont {N.}~\bibnamefont
  {Arkani-Hamed}}, \bibinfo {author} {\bibfnamefont {H.-C.}\ \bibnamefont
  {Cheng}}, \bibinfo {author} {\bibfnamefont {M.~A.}\ \bibnamefont {Luty}}, \
  and\ \bibinfo {author} {\bibfnamefont {S.}~\bibnamefont {Mukohyama}},\ }\href
  {\doibase 10.1088/1126-6708/2004/05/074} {\bibfield  {journal} {\bibinfo
  {journal} {JHEP}\ }\textbf {\bibinfo {volume} {05}},\ \bibinfo {pages} {074}
  (\bibinfo {year} {2004}{\natexlab{a}})},\ \Eprint
  {http://arxiv.org/abs/hep-th/0312099} {arXiv:hep-th/0312099} \BibitemShut
  {NoStop}%
\bibitem [{\citenamefont {Arkani-Hamed}\ \emph
  {et~al.}(2004{\natexlab{b}})\citenamefont {Arkani-Hamed}, \citenamefont
  {Creminelli}, \citenamefont {Mukohyama},\ and\ \citenamefont
  {Zaldarriaga}}]{ArkaniHamed:2003uz}%
  \BibitemOpen
  \bibfield  {author} {\bibinfo {author} {\bibfnamefont {N.}~\bibnamefont
  {Arkani-Hamed}}, \bibinfo {author} {\bibfnamefont {P.}~\bibnamefont
  {Creminelli}}, \bibinfo {author} {\bibfnamefont {S.}~\bibnamefont
  {Mukohyama}}, \ and\ \bibinfo {author} {\bibfnamefont {M.}~\bibnamefont
  {Zaldarriaga}},\ }\href {\doibase 10.1088/1475-7516/2004/04/001} {\bibfield
  {journal} {\bibinfo  {journal} {JCAP}\ }\textbf {\bibinfo {volume} {04}},\
  \bibinfo {pages} {001} (\bibinfo {year} {2004}{\natexlab{b}})},\ \Eprint
  {http://arxiv.org/abs/hep-th/0312100} {arXiv:hep-th/0312100} \BibitemShut
  {NoStop}%
\bibitem [{\citenamefont {Alishahiha}\ \emph {et~al.}(2004)\citenamefont
  {Alishahiha}, \citenamefont {Silverstein},\ and\ \citenamefont
  {Tong}}]{Alishahiha:2004eh}%
  \BibitemOpen
  \bibfield  {author} {\bibinfo {author} {\bibfnamefont {M.}~\bibnamefont
  {Alishahiha}}, \bibinfo {author} {\bibfnamefont {E.}~\bibnamefont
  {Silverstein}}, \ and\ \bibinfo {author} {\bibfnamefont {D.}~\bibnamefont
  {Tong}},\ }\href {\doibase 10.1103/PhysRevD.70.123505} {\bibfield  {journal}
  {\bibinfo  {journal} {Phys. Rev. D}\ }\textbf {\bibinfo {volume} {70}},\
  \bibinfo {pages} {123505} (\bibinfo {year} {2004})},\ \Eprint
  {http://arxiv.org/abs/hep-th/0404084} {arXiv:hep-th/0404084} \BibitemShut
  {NoStop}%
\bibitem [{\citenamefont {Mijic}\ \emph {et~al.}(1986)\citenamefont {Mijic},
  \citenamefont {Morris},\ and\ \citenamefont {Suen}}]{Mijic:1986iv}%
  \BibitemOpen
  \bibfield  {author} {\bibinfo {author} {\bibfnamefont {M.~B.}\ \bibnamefont
  {Mijic}}, \bibinfo {author} {\bibfnamefont {M.~S.}\ \bibnamefont {Morris}}, \
  and\ \bibinfo {author} {\bibfnamefont {W.-M.}\ \bibnamefont {Suen}},\ }\href
  {\doibase 10.1103/PhysRevD.34.2934} {\bibfield  {journal} {\bibinfo
  {journal} {Phys. Rev. D}\ }\textbf {\bibinfo {volume} {34}},\ \bibinfo
  {pages} {2934} (\bibinfo {year} {1986})}\BibitemShut {NoStop}%
\bibitem [{\citenamefont {De~Felice}\ and\ \citenamefont
  {Tsujikawa}(2010)}]{DeFelice:2010aj}%
  \BibitemOpen
  \bibfield  {author} {\bibinfo {author} {\bibfnamefont {A.}~\bibnamefont
  {De~Felice}}\ and\ \bibinfo {author} {\bibfnamefont {S.}~\bibnamefont
  {Tsujikawa}},\ }\href {\doibase 10.12942/lrr-2010-3} {\bibfield  {journal}
  {\bibinfo  {journal} {Living Rev. Rel.}\ }\textbf {\bibinfo {volume} {13}},\
  \bibinfo {pages} {3} (\bibinfo {year} {2010})},\ \Eprint
  {http://arxiv.org/abs/1002.4928} {arXiv:1002.4928 [gr-qc]} \BibitemShut
  {NoStop}%
\bibitem [{\citenamefont {Bamba}\ and\ \citenamefont
  {Odintsov}(2015)}]{Bamba:2015uma}%
  \BibitemOpen
  \bibfield  {author} {\bibinfo {author} {\bibfnamefont {K.}~\bibnamefont
  {Bamba}}\ and\ \bibinfo {author} {\bibfnamefont {S.~D.}\ \bibnamefont
  {Odintsov}},\ }\href {\doibase 10.3390/sym7010220} {\bibfield  {journal}
  {\bibinfo  {journal} {Symmetry}\ }\textbf {\bibinfo {volume} {7}},\ \bibinfo
  {pages} {220} (\bibinfo {year} {2015})},\ \Eprint
  {http://arxiv.org/abs/1503.00442} {arXiv:1503.00442 [hep-th]} \BibitemShut
  {NoStop}%
\bibitem [{\citenamefont {Baumann}\ \emph {et~al.}(2016)\citenamefont
  {Baumann}, \citenamefont {Lee},\ and\ \citenamefont
  {Pimentel}}]{Baumann:2015xxa}%
  \BibitemOpen
  \bibfield  {author} {\bibinfo {author} {\bibfnamefont {D.}~\bibnamefont
  {Baumann}}, \bibinfo {author} {\bibfnamefont {H.}~\bibnamefont {Lee}}, \ and\
  \bibinfo {author} {\bibfnamefont {G.~L.}\ \bibnamefont {Pimentel}},\ }\href
  {\doibase 10.1007/JHEP01(2016)101} {\bibfield  {journal} {\bibinfo  {journal}
  {JHEP}\ }\textbf {\bibinfo {volume} {01}},\ \bibinfo {pages} {101} (\bibinfo
  {year} {2016})},\ \Eprint {http://arxiv.org/abs/1507.07250} {arXiv:1507.07250
  [hep-th]} \BibitemShut {NoStop}%
\bibitem [{\citenamefont {Kobayashi}\ \emph
  {et~al.}(2011{\natexlab{a}})\citenamefont {Kobayashi}, \citenamefont
  {Yamaguchi},\ and\ \citenamefont {Yokoyama}}]{Kobayashi:2011nu}%
  \BibitemOpen
  \bibfield  {author} {\bibinfo {author} {\bibfnamefont {T.}~\bibnamefont
  {Kobayashi}}, \bibinfo {author} {\bibfnamefont {M.}~\bibnamefont
  {Yamaguchi}}, \ and\ \bibinfo {author} {\bibfnamefont {J.}~\bibnamefont
  {Yokoyama}},\ }\href {\doibase 10.1143/PTP.126.511} {\bibfield  {journal}
  {\bibinfo  {journal} {Prog. Theor. Phys.}\ }\textbf {\bibinfo {volume}
  {126}},\ \bibinfo {pages} {511} (\bibinfo {year} {2011}{\natexlab{a}})},\
  \Eprint {http://arxiv.org/abs/1105.5723} {arXiv:1105.5723 [hep-th]}
  \BibitemShut {NoStop}%
\bibitem [{\citenamefont {Chen}\ \emph {et~al.}(2007)\citenamefont {Chen},
  \citenamefont {Huang}, \citenamefont {Kachru},\ and\ \citenamefont
  {Shiu}}]{Chen:2006nt}%
  \BibitemOpen
  \bibfield  {author} {\bibinfo {author} {\bibfnamefont {X.}~\bibnamefont
  {Chen}}, \bibinfo {author} {\bibfnamefont {M.-x.}\ \bibnamefont {Huang}},
  \bibinfo {author} {\bibfnamefont {S.}~\bibnamefont {Kachru}}, \ and\ \bibinfo
  {author} {\bibfnamefont {G.}~\bibnamefont {Shiu}},\ }\href {\doibase
  10.1088/1475-7516/2007/01/002} {\bibfield  {journal} {\bibinfo  {journal}
  {JCAP}\ }\textbf {\bibinfo {volume} {01}},\ \bibinfo {pages} {002} (\bibinfo
  {year} {2007})},\ \Eprint {http://arxiv.org/abs/hep-th/0605045}
  {arXiv:hep-th/0605045} \BibitemShut {NoStop}%
\bibitem [{\citenamefont {Chen}\ \emph {et~al.}(2009)\citenamefont {Chen},
  \citenamefont {Hu}, \citenamefont {Huang}, \citenamefont {Shiu},\ and\
  \citenamefont {Wang}}]{Chen:2009bc}%
  \BibitemOpen
  \bibfield  {author} {\bibinfo {author} {\bibfnamefont {X.}~\bibnamefont
  {Chen}}, \bibinfo {author} {\bibfnamefont {B.}~\bibnamefont {Hu}}, \bibinfo
  {author} {\bibfnamefont {M.-x.}\ \bibnamefont {Huang}}, \bibinfo {author}
  {\bibfnamefont {G.}~\bibnamefont {Shiu}}, \ and\ \bibinfo {author}
  {\bibfnamefont {Y.}~\bibnamefont {Wang}},\ }\href {\doibase
  10.1088/1475-7516/2009/08/008} {\bibfield  {journal} {\bibinfo  {journal}
  {JCAP}\ }\textbf {\bibinfo {volume} {08}},\ \bibinfo {pages} {008} (\bibinfo
  {year} {2009})},\ \Eprint {http://arxiv.org/abs/0905.3494} {arXiv:0905.3494
  [astro-ph.CO]} \BibitemShut {NoStop}%
\bibitem [{\citenamefont {Chen}\ \emph {et~al.}(2013)\citenamefont {Chen},
  \citenamefont {Firouzjahi}, \citenamefont {Namjoo},\ and\ \citenamefont
  {Sasaki}}]{Chen:2013aj}%
  \BibitemOpen
  \bibfield  {author} {\bibinfo {author} {\bibfnamefont {X.}~\bibnamefont
  {Chen}}, \bibinfo {author} {\bibfnamefont {H.}~\bibnamefont {Firouzjahi}},
  \bibinfo {author} {\bibfnamefont {M.~H.}\ \bibnamefont {Namjoo}}, \ and\
  \bibinfo {author} {\bibfnamefont {M.}~\bibnamefont {Sasaki}},\ }\href
  {\doibase 10.1209/0295-5075/102/59001} {\bibfield  {journal} {\bibinfo
  {journal} {EPL}\ }\textbf {\bibinfo {volume} {102}},\ \bibinfo {pages}
  {59001} (\bibinfo {year} {2013})},\ \Eprint {http://arxiv.org/abs/1301.5699}
  {arXiv:1301.5699 [hep-th]} \BibitemShut {NoStop}%
\bibitem [{\citenamefont {Seery}\ and\ \citenamefont
  {Lidsey}(2005)}]{Seery:2005wm}%
  \BibitemOpen
  \bibfield  {author} {\bibinfo {author} {\bibfnamefont {D.}~\bibnamefont
  {Seery}}\ and\ \bibinfo {author} {\bibfnamefont {J.~E.}\ \bibnamefont
  {Lidsey}},\ }\href {\doibase 10.1088/1475-7516/2005/06/003} {\bibfield
  {journal} {\bibinfo  {journal} {JCAP}\ }\textbf {\bibinfo {volume} {06}},\
  \bibinfo {pages} {003} (\bibinfo {year} {2005})},\ \Eprint
  {http://arxiv.org/abs/astro-ph/0503692} {arXiv:astro-ph/0503692} \BibitemShut
  {NoStop}%
\bibitem [{\citenamefont {Kobayashi}\ \emph
  {et~al.}(2011{\natexlab{b}})\citenamefont {Kobayashi}, \citenamefont
  {Yamaguchi},\ and\ \citenamefont {Yokoyama}}]{Kobayashi:2011pc}%
  \BibitemOpen
  \bibfield  {author} {\bibinfo {author} {\bibfnamefont {T.}~\bibnamefont
  {Kobayashi}}, \bibinfo {author} {\bibfnamefont {M.}~\bibnamefont
  {Yamaguchi}}, \ and\ \bibinfo {author} {\bibfnamefont {J.}~\bibnamefont
  {Yokoyama}},\ }\href {\doibase 10.1103/PhysRevD.83.103524} {\bibfield
  {journal} {\bibinfo  {journal} {Phys. Rev. D}\ }\textbf {\bibinfo {volume}
  {83}},\ \bibinfo {pages} {103524} (\bibinfo {year} {2011}{\natexlab{b}})},\
  \Eprint {http://arxiv.org/abs/1103.1740} {arXiv:1103.1740 [hep-th]}
  \BibitemShut {NoStop}%
\bibitem [{\citenamefont {Baumann}\ and\ \citenamefont
  {Green}(2012)}]{Baumann:2011nk}%
  \BibitemOpen
  \bibfield  {author} {\bibinfo {author} {\bibfnamefont {D.}~\bibnamefont
  {Baumann}}\ and\ \bibinfo {author} {\bibfnamefont {D.}~\bibnamefont
  {Green}},\ }\href {\doibase 10.1103/PhysRevD.85.103520} {\bibfield  {journal}
  {\bibinfo  {journal} {Phys. Rev. D}\ }\textbf {\bibinfo {volume} {85}},\
  \bibinfo {pages} {103520} (\bibinfo {year} {2012})},\ \Eprint
  {http://arxiv.org/abs/1109.0292} {arXiv:1109.0292 [hep-th]} \BibitemShut
  {NoStop}%
\bibitem [{\citenamefont {Baumann}\ and\ \citenamefont
  {Green}(2011)}]{Baumann:2011su}%
  \BibitemOpen
  \bibfield  {author} {\bibinfo {author} {\bibfnamefont {D.}~\bibnamefont
  {Baumann}}\ and\ \bibinfo {author} {\bibfnamefont {D.}~\bibnamefont
  {Green}},\ }\href {\doibase 10.1088/1475-7516/2011/09/014} {\bibfield
  {journal} {\bibinfo  {journal} {JCAP}\ }\textbf {\bibinfo {volume} {09}},\
  \bibinfo {pages} {014} (\bibinfo {year} {2011})},\ \Eprint
  {http://arxiv.org/abs/1102.5343} {arXiv:1102.5343 [hep-th]} \BibitemShut
  {NoStop}%
\bibitem [{\citenamefont {Weinberg}(2005)}]{Weinberg:2005vy}%
  \BibitemOpen
  \bibfield  {author} {\bibinfo {author} {\bibfnamefont {S.}~\bibnamefont
  {Weinberg}},\ }\href {\doibase 10.1103/PhysRevD.72.043514} {\bibfield
  {journal} {\bibinfo  {journal} {Phys. Rev. D}\ }\textbf {\bibinfo {volume}
  {72}},\ \bibinfo {pages} {043514} (\bibinfo {year} {2005})},\ \Eprint
  {http://arxiv.org/abs/hep-th/0506236} {arXiv:hep-th/0506236} \BibitemShut
  {NoStop}%
\bibitem [{\citenamefont {Weinberg}(2006)}]{Weinberg:2006ac}%
  \BibitemOpen
  \bibfield  {author} {\bibinfo {author} {\bibfnamefont {S.}~\bibnamefont
  {Weinberg}},\ }\href {\doibase 10.1103/PhysRevD.74.023508} {\bibfield
  {journal} {\bibinfo  {journal} {Phys. Rev. D}\ }\textbf {\bibinfo {volume}
  {74}},\ \bibinfo {pages} {023508} (\bibinfo {year} {2006})},\ \Eprint
  {http://arxiv.org/abs/hep-th/0605244} {arXiv:hep-th/0605244} \BibitemShut
  {NoStop}%
\bibitem [{\citenamefont {Sloth}(2006)}]{Sloth:2006az}%
  \BibitemOpen
  \bibfield  {author} {\bibinfo {author} {\bibfnamefont {M.~S.}\ \bibnamefont
  {Sloth}},\ }\href {\doibase 10.1016/j.nuclphysb.2006.04.029} {\bibfield
  {journal} {\bibinfo  {journal} {Nucl. Phys. B}\ }\textbf {\bibinfo {volume}
  {748}},\ \bibinfo {pages} {149} (\bibinfo {year} {2006})},\ \Eprint
  {http://arxiv.org/abs/astro-ph/0604488} {arXiv:astro-ph/0604488} \BibitemShut
  {NoStop}%
\bibitem [{\citenamefont {Sloth}(2007)}]{Sloth:2006nu}%
  \BibitemOpen
  \bibfield  {author} {\bibinfo {author} {\bibfnamefont {M.~S.}\ \bibnamefont
  {Sloth}},\ }\href {\doibase 10.1016/j.nuclphysb.2007.04.012} {\bibfield
  {journal} {\bibinfo  {journal} {Nucl. Phys. B}\ }\textbf {\bibinfo {volume}
  {775}},\ \bibinfo {pages} {78} (\bibinfo {year} {2007})},\ \Eprint
  {http://arxiv.org/abs/hep-th/0612138} {arXiv:hep-th/0612138} \BibitemShut
  {NoStop}%
\bibitem [{\citenamefont {Seery}(2007)}]{Seery:2007we}%
  \BibitemOpen
  \bibfield  {author} {\bibinfo {author} {\bibfnamefont {D.}~\bibnamefont
  {Seery}},\ }\href {\doibase 10.1088/1475-7516/2007/11/025} {\bibfield
  {journal} {\bibinfo  {journal} {JCAP}\ }\textbf {\bibinfo {volume} {11}},\
  \bibinfo {pages} {025} (\bibinfo {year} {2007})},\ \Eprint
  {http://arxiv.org/abs/0707.3377} {arXiv:0707.3377 [astro-ph]} \BibitemShut
  {NoStop}%
\bibitem [{\citenamefont {Seery}(2008)}]{Seery:2007wf}%
  \BibitemOpen
  \bibfield  {author} {\bibinfo {author} {\bibfnamefont {D.}~\bibnamefont
  {Seery}},\ }\href {\doibase 10.1088/1475-7516/2008/02/006} {\bibfield
  {journal} {\bibinfo  {journal} {JCAP}\ }\textbf {\bibinfo {volume} {02}},\
  \bibinfo {pages} {006} (\bibinfo {year} {2008})},\ \Eprint
  {http://arxiv.org/abs/0707.3378} {arXiv:0707.3378 [astro-ph]} \BibitemShut
  {NoStop}%
\bibitem [{\citenamefont {Byrnes}\ \emph {et~al.}(2007)\citenamefont {Byrnes},
  \citenamefont {Koyama}, \citenamefont {Sasaki},\ and\ \citenamefont
  {Wands}}]{Byrnes:2007tm}%
  \BibitemOpen
  \bibfield  {author} {\bibinfo {author} {\bibfnamefont {C.~T.}\ \bibnamefont
  {Byrnes}}, \bibinfo {author} {\bibfnamefont {K.}~\bibnamefont {Koyama}},
  \bibinfo {author} {\bibfnamefont {M.}~\bibnamefont {Sasaki}}, \ and\ \bibinfo
  {author} {\bibfnamefont {D.}~\bibnamefont {Wands}},\ }\href {\doibase
  10.1088/1475-7516/2007/11/027} {\bibfield  {journal} {\bibinfo  {journal}
  {JCAP}\ }\textbf {\bibinfo {volume} {11}},\ \bibinfo {pages} {027} (\bibinfo
  {year} {2007})},\ \Eprint {http://arxiv.org/abs/0705.4096} {arXiv:0705.4096
  [hep-th]} \BibitemShut {NoStop}%
\bibitem [{\citenamefont {Dimastrogiovanni}\ and\ \citenamefont
  {Bartolo}(2008)}]{Dimastrogiovanni:2008af}%
  \BibitemOpen
  \bibfield  {author} {\bibinfo {author} {\bibfnamefont {E.}~\bibnamefont
  {Dimastrogiovanni}}\ and\ \bibinfo {author} {\bibfnamefont {N.}~\bibnamefont
  {Bartolo}},\ }\href {\doibase 10.1088/1475-7516/2008/11/016} {\bibfield
  {journal} {\bibinfo  {journal} {JCAP}\ }\textbf {\bibinfo {volume} {11}},\
  \bibinfo {pages} {016} (\bibinfo {year} {2008})},\ \Eprint
  {http://arxiv.org/abs/0807.2790} {arXiv:0807.2790 [astro-ph]} \BibitemShut
  {NoStop}%
\bibitem [{\citenamefont {Senatore}\ and\ \citenamefont
  {Zaldarriaga}(2010)}]{Senatore:2009cf}%
  \BibitemOpen
  \bibfield  {author} {\bibinfo {author} {\bibfnamefont {L.}~\bibnamefont
  {Senatore}}\ and\ \bibinfo {author} {\bibfnamefont {M.}~\bibnamefont
  {Zaldarriaga}},\ }\href {\doibase 10.1007/JHEP12(2010)008} {\bibfield
  {journal} {\bibinfo  {journal} {JHEP}\ }\textbf {\bibinfo {volume} {12}},\
  \bibinfo {pages} {008} (\bibinfo {year} {2010})},\ \Eprint
  {http://arxiv.org/abs/0912.2734} {arXiv:0912.2734 [hep-th]} \BibitemShut
  {NoStop}%
\bibitem [{\citenamefont {Bartolo}\ \emph {et~al.}(2010)\citenamefont
  {Bartolo}, \citenamefont {Dimastrogiovanni},\ and\ \citenamefont
  {Vallinotto}}]{Bartolo:2010bu}%
  \BibitemOpen
  \bibfield  {author} {\bibinfo {author} {\bibfnamefont {N.}~\bibnamefont
  {Bartolo}}, \bibinfo {author} {\bibfnamefont {E.}~\bibnamefont
  {Dimastrogiovanni}}, \ and\ \bibinfo {author} {\bibfnamefont
  {A.}~\bibnamefont {Vallinotto}},\ }\href {\doibase
  10.1088/1475-7516/2010/11/003} {\bibfield  {journal} {\bibinfo  {journal}
  {JCAP}\ }\textbf {\bibinfo {volume} {11}},\ \bibinfo {pages} {003} (\bibinfo
  {year} {2010})},\ \Eprint {http://arxiv.org/abs/1006.0196} {arXiv:1006.0196
  [astro-ph.CO]} \BibitemShut {NoStop}%
\bibitem [{\citenamefont {Giddings}\ and\ \citenamefont
  {Sloth}(2011)}]{Giddings:2010nc}%
  \BibitemOpen
  \bibfield  {author} {\bibinfo {author} {\bibfnamefont {S.~B.}\ \bibnamefont
  {Giddings}}\ and\ \bibinfo {author} {\bibfnamefont {M.~S.}\ \bibnamefont
  {Sloth}},\ }\href {\doibase 10.1088/1475-7516/2011/01/023} {\bibfield
  {journal} {\bibinfo  {journal} {JCAP}\ }\textbf {\bibinfo {volume} {01}},\
  \bibinfo {pages} {023} (\bibinfo {year} {2011})},\ \Eprint
  {http://arxiv.org/abs/1005.1056} {arXiv:1005.1056 [hep-th]} \BibitemShut
  {NoStop}%
\bibitem [{\citenamefont {Leblond}\ and\ \citenamefont
  {Shandera}(2008)}]{Leblond:2008gg}%
  \BibitemOpen
  \bibfield  {author} {\bibinfo {author} {\bibfnamefont {L.}~\bibnamefont
  {Leblond}}\ and\ \bibinfo {author} {\bibfnamefont {S.}~\bibnamefont
  {Shandera}},\ }\href {\doibase 10.1088/1475-7516/2008/08/007} {\bibfield
  {journal} {\bibinfo  {journal} {JCAP}\ }\textbf {\bibinfo {volume} {08}},\
  \bibinfo {pages} {007} (\bibinfo {year} {2008})},\ \Eprint
  {http://arxiv.org/abs/0802.2290} {arXiv:0802.2290 [hep-th]} \BibitemShut
  {NoStop}%
\bibitem [{\citenamefont {Assassi}\ \emph {et~al.}(2013)\citenamefont
  {Assassi}, \citenamefont {Baumann},\ and\ \citenamefont
  {Green}}]{Assassi:2012et}%
  \BibitemOpen
  \bibfield  {author} {\bibinfo {author} {\bibfnamefont {V.}~\bibnamefont
  {Assassi}}, \bibinfo {author} {\bibfnamefont {D.}~\bibnamefont {Baumann}}, \
  and\ \bibinfo {author} {\bibfnamefont {D.}~\bibnamefont {Green}},\ }\href
  {\doibase 10.1007/JHEP02(2013)151} {\bibfield  {journal} {\bibinfo  {journal}
  {JHEP}\ }\textbf {\bibinfo {volume} {02}},\ \bibinfo {pages} {151} (\bibinfo
  {year} {2013})},\ \Eprint {http://arxiv.org/abs/1210.7792} {arXiv:1210.7792
  [hep-th]} \BibitemShut {NoStop}%
\bibitem [{\citenamefont {Melville}\ and\ \citenamefont
  {Pajer}(2021)}]{Melville:2021lst}%
  \BibitemOpen
  \bibfield  {author} {\bibinfo {author} {\bibfnamefont {S.}~\bibnamefont
  {Melville}}\ and\ \bibinfo {author} {\bibfnamefont {E.}~\bibnamefont
  {Pajer}},\ }\href {\doibase 10.1007/JHEP05(2021)249} {\bibfield  {journal}
  {\bibinfo  {journal} {JHEP}\ }\textbf {\bibinfo {volume} {05}},\ \bibinfo
  {pages} {249} (\bibinfo {year} {2021})},\ \Eprint
  {http://arxiv.org/abs/2103.09832} {arXiv:2103.09832 [hep-th]} \BibitemShut
  {NoStop}%
\bibitem [{\citenamefont {Tanaka}\ and\ \citenamefont
  {Urakawa}(2013)}]{Tanaka:2013caa}%
  \BibitemOpen
  \bibfield  {author} {\bibinfo {author} {\bibfnamefont {T.}~\bibnamefont
  {Tanaka}}\ and\ \bibinfo {author} {\bibfnamefont {Y.}~\bibnamefont
  {Urakawa}},\ }\href {\doibase 10.1088/0264-9381/30/23/233001} {\bibfield
  {journal} {\bibinfo  {journal} {Class. Quant. Grav.}\ }\textbf {\bibinfo
  {volume} {30}},\ \bibinfo {pages} {233001} (\bibinfo {year} {2013})},\
  \Eprint {http://arxiv.org/abs/1306.4461} {arXiv:1306.4461 [hep-th]}
  \BibitemShut {NoStop}%
\bibitem [{\citenamefont {Urakawa}\ and\ \citenamefont
  {Tanaka}(2011)}]{Urakawa:2010kr}%
  \BibitemOpen
  \bibfield  {author} {\bibinfo {author} {\bibfnamefont {Y.}~\bibnamefont
  {Urakawa}}\ and\ \bibinfo {author} {\bibfnamefont {T.}~\bibnamefont
  {Tanaka}},\ }\href {\doibase 10.1143/PTP.125.1067} {\bibfield  {journal}
  {\bibinfo  {journal} {Prog. Theor. Phys.}\ }\textbf {\bibinfo {volume}
  {125}},\ \bibinfo {pages} {1067} (\bibinfo {year} {2011})},\ \Eprint
  {http://arxiv.org/abs/1009.2947} {arXiv:1009.2947 [hep-th]} \BibitemShut
  {NoStop}%
\bibitem [{\citenamefont {Chen}\ \emph {et~al.}(2016)\citenamefont {Chen},
  \citenamefont {Wang},\ and\ \citenamefont {Xianyu}}]{Chen:2016nrs}%
  \BibitemOpen
  \bibfield  {author} {\bibinfo {author} {\bibfnamefont {X.}~\bibnamefont
  {Chen}}, \bibinfo {author} {\bibfnamefont {Y.}~\bibnamefont {Wang}}, \ and\
  \bibinfo {author} {\bibfnamefont {Z.-Z.}\ \bibnamefont {Xianyu}},\ }\href
  {\doibase 10.1007/JHEP08(2016)051} {\bibfield  {journal} {\bibinfo  {journal}
  {JHEP}\ }\textbf {\bibinfo {volume} {08}},\ \bibinfo {pages} {051} (\bibinfo
  {year} {2016})},\ \Eprint {http://arxiv.org/abs/1604.07841} {arXiv:1604.07841
  [hep-th]} \BibitemShut {NoStop}%
\bibitem [{\citenamefont {Chen}\ \emph {et~al.}(2017)\citenamefont {Chen},
  \citenamefont {Wang},\ and\ \citenamefont {Xianyu}}]{Chen:2016uwp}%
  \BibitemOpen
  \bibfield  {author} {\bibinfo {author} {\bibfnamefont {X.}~\bibnamefont
  {Chen}}, \bibinfo {author} {\bibfnamefont {Y.}~\bibnamefont {Wang}}, \ and\
  \bibinfo {author} {\bibfnamefont {Z.-Z.}\ \bibnamefont {Xianyu}},\ }\href
  {\doibase 10.1103/PhysRevLett.118.261302} {\bibfield  {journal} {\bibinfo
  {journal} {Phys. Rev. Lett.}\ }\textbf {\bibinfo {volume} {118}},\ \bibinfo
  {pages} {261302} (\bibinfo {year} {2017})},\ \Eprint
  {http://arxiv.org/abs/1610.06597} {arXiv:1610.06597 [hep-th]} \BibitemShut
  {NoStop}%
\bibitem [{\citenamefont {Wu}\ and\ \citenamefont
  {Yokoyama}(2018)}]{Wu:2017lnh}%
  \BibitemOpen
  \bibfield  {author} {\bibinfo {author} {\bibfnamefont {Y.-P.}\ \bibnamefont
  {Wu}}\ and\ \bibinfo {author} {\bibfnamefont {J.}~\bibnamefont {Yokoyama}},\
  }\href {\doibase 10.1088/1475-7516/2018/05/009} {\bibfield  {journal}
  {\bibinfo  {journal} {JCAP}\ }\textbf {\bibinfo {volume} {05}},\ \bibinfo
  {pages} {009} (\bibinfo {year} {2018})},\ \Eprint
  {http://arxiv.org/abs/1704.05026} {arXiv:1704.05026 [hep-th]} \BibitemShut
  {NoStop}%
\bibitem [{\citenamefont {Sasaki}(1986)}]{Sasaki:1986hm}%
  \BibitemOpen
  \bibfield  {author} {\bibinfo {author} {\bibfnamefont {M.}~\bibnamefont
  {Sasaki}},\ }\href {\doibase 10.1143/PTP.76.1036} {\bibfield  {journal}
  {\bibinfo  {journal} {Prog. Theor. Phys.}\ }\textbf {\bibinfo {volume}
  {76}},\ \bibinfo {pages} {1036} (\bibinfo {year} {1986})}\BibitemShut
  {NoStop}%
\bibitem [{\citenamefont {Cheung}\ \emph {et~al.}(2008)\citenamefont {Cheung},
  \citenamefont {Creminelli}, \citenamefont {Fitzpatrick}, \citenamefont
  {Kaplan},\ and\ \citenamefont {Senatore}}]{Cheung:2007st}%
  \BibitemOpen
  \bibfield  {author} {\bibinfo {author} {\bibfnamefont {C.}~\bibnamefont
  {Cheung}}, \bibinfo {author} {\bibfnamefont {P.}~\bibnamefont {Creminelli}},
  \bibinfo {author} {\bibfnamefont {A.~L.}\ \bibnamefont {Fitzpatrick}},
  \bibinfo {author} {\bibfnamefont {J.}~\bibnamefont {Kaplan}}, \ and\ \bibinfo
  {author} {\bibfnamefont {L.}~\bibnamefont {Senatore}},\ }\href {\doibase
  10.1088/1126-6708/2008/03/014} {\bibfield  {journal} {\bibinfo  {journal}
  {JHEP}\ }\textbf {\bibinfo {volume} {03}},\ \bibinfo {pages} {014} (\bibinfo
  {year} {2008})},\ \Eprint {http://arxiv.org/abs/0709.0293} {arXiv:0709.0293
  [hep-th]} \BibitemShut {NoStop}%
\bibitem [{\citenamefont {Adshead}\ \emph {et~al.}(2009)\citenamefont
  {Adshead}, \citenamefont {Easther},\ and\ \citenamefont
  {Lim}}]{Adshead:2008gk}%
  \BibitemOpen
  \bibfield  {author} {\bibinfo {author} {\bibfnamefont {P.}~\bibnamefont
  {Adshead}}, \bibinfo {author} {\bibfnamefont {R.}~\bibnamefont {Easther}}, \
  and\ \bibinfo {author} {\bibfnamefont {E.~A.}\ \bibnamefont {Lim}},\ }\href
  {\doibase 10.1103/PhysRevD.79.063504} {\bibfield  {journal} {\bibinfo
  {journal} {Phys. Rev. D}\ }\textbf {\bibinfo {volume} {79}},\ \bibinfo
  {pages} {063504} (\bibinfo {year} {2009})},\ \Eprint
  {http://arxiv.org/abs/0809.4008} {arXiv:0809.4008 [hep-th]} \BibitemShut
  {NoStop}%
\bibitem [{Note1()}]{Note1}%
  \BibitemOpen
  \bibinfo {note} {Here one may recall Einstein's saying that the eternal
  mystery of the world is its comprehensibility.}\BibitemShut {Stop}%
\bibitem [{\citenamefont {Baumann}\ \emph {et~al.}(2011)\citenamefont
  {Baumann}, \citenamefont {Senatore},\ and\ \citenamefont
  {Zaldarriaga}}]{Baumann:2011dt}%
  \BibitemOpen
  \bibfield  {author} {\bibinfo {author} {\bibfnamefont {D.}~\bibnamefont
  {Baumann}}, \bibinfo {author} {\bibfnamefont {L.}~\bibnamefont {Senatore}}, \
  and\ \bibinfo {author} {\bibfnamefont {M.}~\bibnamefont {Zaldarriaga}},\
  }\href {\doibase 10.1088/1475-7516/2011/05/004} {\bibfield  {journal}
  {\bibinfo  {journal} {JCAP}\ }\textbf {\bibinfo {volume} {05}},\ \bibinfo
  {pages} {004} (\bibinfo {year} {2011})},\ \Eprint
  {http://arxiv.org/abs/1101.3320} {arXiv:1101.3320 [hep-th]} \BibitemShut
  {NoStop}%
\bibitem [{\citenamefont {Stewart}\ and\ \citenamefont
  {Lyth}(1993)}]{Stewart:1993bc}%
  \BibitemOpen
  \bibfield  {author} {\bibinfo {author} {\bibfnamefont {E.~D.}\ \bibnamefont
  {Stewart}}\ and\ \bibinfo {author} {\bibfnamefont {D.~H.}\ \bibnamefont
  {Lyth}},\ }\href {\doibase 10.1016/0370-2693(93)90379-V} {\bibfield
  {journal} {\bibinfo  {journal} {Phys. Lett. B}\ }\textbf {\bibinfo {volume}
  {302}},\ \bibinfo {pages} {171} (\bibinfo {year} {1993})},\ \Eprint
  {http://arxiv.org/abs/gr-qc/9302019} {arXiv:gr-qc/9302019} \BibitemShut
  {NoStop}%
\bibitem [{\citenamefont {Jeong}\ \emph {et~al.}(2014)\citenamefont {Jeong},
  \citenamefont {Pradler}, \citenamefont {Chluba},\ and\ \citenamefont
  {Kamionkowski}}]{Jeong:2014gna}%
  \BibitemOpen
  \bibfield  {author} {\bibinfo {author} {\bibfnamefont {D.}~\bibnamefont
  {Jeong}}, \bibinfo {author} {\bibfnamefont {J.}~\bibnamefont {Pradler}},
  \bibinfo {author} {\bibfnamefont {J.}~\bibnamefont {Chluba}}, \ and\ \bibinfo
  {author} {\bibfnamefont {M.}~\bibnamefont {Kamionkowski}},\ }\href {\doibase
  10.1103/PhysRevLett.113.061301} {\bibfield  {journal} {\bibinfo  {journal}
  {Phys. Rev. Lett.}\ }\textbf {\bibinfo {volume} {113}},\ \bibinfo {pages}
  {061301} (\bibinfo {year} {2014})},\ \Eprint {http://arxiv.org/abs/1403.3697}
  {arXiv:1403.3697 [astro-ph.CO]} \BibitemShut {NoStop}%
\bibitem [{\citenamefont {Nakama}\ \emph {et~al.}(2014)\citenamefont {Nakama},
  \citenamefont {Suyama},\ and\ \citenamefont {Yokoyama}}]{Nakama:2014vla}%
  \BibitemOpen
  \bibfield  {author} {\bibinfo {author} {\bibfnamefont {T.}~\bibnamefont
  {Nakama}}, \bibinfo {author} {\bibfnamefont {T.}~\bibnamefont {Suyama}}, \
  and\ \bibinfo {author} {\bibfnamefont {J.}~\bibnamefont {Yokoyama}},\ }\href
  {\doibase 10.1103/PhysRevLett.113.061302} {\bibfield  {journal} {\bibinfo
  {journal} {Phys. Rev. Lett.}\ }\textbf {\bibinfo {volume} {113}},\ \bibinfo
  {pages} {061302} (\bibinfo {year} {2014})},\ \Eprint
  {http://arxiv.org/abs/1403.5407} {arXiv:1403.5407 [astro-ph.CO]} \BibitemShut
  {NoStop}%
\bibitem [{\citenamefont {{Zel'dovich}}\ and\ \citenamefont
  {{Novikov}}(1967)}]{1967SvA....10..602Z}%
  \BibitemOpen
  \bibfield  {author} {\bibinfo {author} {\bibfnamefont {Y.~B.}\ \bibnamefont
  {{Zel'dovich}}}\ and\ \bibinfo {author} {\bibfnamefont {I.~D.}\ \bibnamefont
  {{Novikov}}},\ }\href@noop {} {\bibfield  {journal} {\bibinfo  {journal}
  {Sov. Astron.}\ }\textbf {\bibinfo {volume} {10}},\ \bibinfo {pages} {602}
  (\bibinfo {year} {1967})}\BibitemShut {NoStop}%
\bibitem [{\citenamefont {Hawking}(1974)}]{Hawking:1974rv}%
  \BibitemOpen
  \bibfield  {author} {\bibinfo {author} {\bibfnamefont {S.~W.}\ \bibnamefont
  {Hawking}},\ }\href {\doibase 10.1038/248030a0} {\bibfield  {journal}
  {\bibinfo  {journal} {Nature}\ }\textbf {\bibinfo {volume} {248}},\ \bibinfo
  {pages} {30} (\bibinfo {year} {1974})}\BibitemShut {NoStop}%
%%CITATION = NATUA,248,30;%%
\bibitem [{\citenamefont {Carr}\ \emph {et~al.}(2010)\citenamefont {Carr},
  \citenamefont {Kohri}, \citenamefont {Sendouda},\ and\ \citenamefont
  {Yokoyama}}]{Carr:2009jm}%
  \BibitemOpen
  \bibfield  {author} {\bibinfo {author} {\bibfnamefont {B.~J.}\ \bibnamefont
  {Carr}}, \bibinfo {author} {\bibfnamefont {K.}~\bibnamefont {Kohri}},
  \bibinfo {author} {\bibfnamefont {Y.}~\bibnamefont {Sendouda}}, \ and\
  \bibinfo {author} {\bibfnamefont {J.}~\bibnamefont {Yokoyama}},\ }\href
  {\doibase 10.1103/PhysRevD.81.104019} {\bibfield  {journal} {\bibinfo
  {journal} {Phys. Rev.}\ }\textbf {\bibinfo {volume} {D81}},\ \bibinfo {pages}
  {104019} (\bibinfo {year} {2010})},\ \Eprint {http://arxiv.org/abs/0912.5297}
  {arXiv:0912.5297 [astro-ph.CO]} \BibitemShut {NoStop}%
%%CITATION = ARXIV:0912.5297;%%
\bibitem [{\citenamefont {Carr}\ \emph {et~al.}(2021)\citenamefont {Carr},
  \citenamefont {Kohri}, \citenamefont {Sendouda},\ and\ \citenamefont
  {Yokoyama}}]{Carr:2020gox}%
  \BibitemOpen
  \bibfield  {author} {\bibinfo {author} {\bibfnamefont {B.}~\bibnamefont
  {Carr}}, \bibinfo {author} {\bibfnamefont {K.}~\bibnamefont {Kohri}},
  \bibinfo {author} {\bibfnamefont {Y.}~\bibnamefont {Sendouda}}, \ and\
  \bibinfo {author} {\bibfnamefont {J.}~\bibnamefont {Yokoyama}},\ }\href
  {\doibase 10.1088/1361-6633/ac1e31} {\bibfield  {journal} {\bibinfo
  {journal} {Rept. Prog. Phys.}\ }\textbf {\bibinfo {volume} {84}},\ \bibinfo
  {pages} {116902} (\bibinfo {year} {2021})},\ \Eprint
  {http://arxiv.org/abs/2002.12778} {arXiv:2002.12778 [astro-ph.CO]}
  \BibitemShut {NoStop}%
\bibitem [{\citenamefont {Baumann}\ \emph {et~al.}(2007)\citenamefont
  {Baumann}, \citenamefont {Steinhardt}, \citenamefont {Takahashi},\ and\
  \citenamefont {Ichiki}}]{Baumann:2007zm}%
  \BibitemOpen
  \bibfield  {author} {\bibinfo {author} {\bibfnamefont {D.}~\bibnamefont
  {Baumann}}, \bibinfo {author} {\bibfnamefont {P.~J.}\ \bibnamefont
  {Steinhardt}}, \bibinfo {author} {\bibfnamefont {K.}~\bibnamefont
  {Takahashi}}, \ and\ \bibinfo {author} {\bibfnamefont {K.}~\bibnamefont
  {Ichiki}},\ }\href {\doibase 10.1103/PhysRevD.76.084019} {\bibfield
  {journal} {\bibinfo  {journal} {Phys. Rev. D}\ }\textbf {\bibinfo {volume}
  {76}},\ \bibinfo {pages} {084019} (\bibinfo {year} {2007})},\ \Eprint
  {http://arxiv.org/abs/hep-th/0703290} {arXiv:hep-th/0703290} \BibitemShut
  {NoStop}%
\bibitem [{\citenamefont {Assadullahi}\ and\ \citenamefont
  {Wands}(2010)}]{Assadullahi:2009jc}%
  \BibitemOpen
  \bibfield  {author} {\bibinfo {author} {\bibfnamefont {H.}~\bibnamefont
  {Assadullahi}}\ and\ \bibinfo {author} {\bibfnamefont {D.}~\bibnamefont
  {Wands}},\ }\href {\doibase 10.1103/PhysRevD.81.023527} {\bibfield  {journal}
  {\bibinfo  {journal} {Phys. Rev. D}\ }\textbf {\bibinfo {volume} {81}},\
  \bibinfo {pages} {023527} (\bibinfo {year} {2010})},\ \Eprint
  {http://arxiv.org/abs/0907.4073} {arXiv:0907.4073 [astro-ph.CO]} \BibitemShut
  {NoStop}%
\bibitem [{\citenamefont {Saito}\ and\ \citenamefont
  {Yokoyama}(2009)}]{Saito:2008jc}%
  \BibitemOpen
  \bibfield  {author} {\bibinfo {author} {\bibfnamefont {R.}~\bibnamefont
  {Saito}}\ and\ \bibinfo {author} {\bibfnamefont {J.}~\bibnamefont
  {Yokoyama}},\ }\href {\doibase 10.1103/PhysRevLett.102.161101} {\bibfield
  {journal} {\bibinfo  {journal} {Phys. Rev. Lett.}\ }\textbf {\bibinfo
  {volume} {102}},\ \bibinfo {pages} {161101} (\bibinfo {year} {2009})},\
  \bibinfo {note} {[Erratum: Phys.Rev.Lett. 107, 069901 (2011)]},\ \Eprint
  {http://arxiv.org/abs/0812.4339} {arXiv:0812.4339 [astro-ph]} \BibitemShut
  {NoStop}%
\bibitem [{\citenamefont {Saito}\ and\ \citenamefont
  {Yokoyama}(2010)}]{Saito:2009jt}%
  \BibitemOpen
  \bibfield  {author} {\bibinfo {author} {\bibfnamefont {R.}~\bibnamefont
  {Saito}}\ and\ \bibinfo {author} {\bibfnamefont {J.}~\bibnamefont
  {Yokoyama}},\ }\href {\doibase 10.1143/PTP.126.351} {\bibfield  {journal}
  {\bibinfo  {journal} {Prog. Theor. Phys.}\ }\textbf {\bibinfo {volume}
  {123}},\ \bibinfo {pages} {867} (\bibinfo {year} {2010})},\ \bibinfo {note}
  {[Erratum: Prog.Theor.Phys. 126, 351--352 (2011)]},\ \Eprint
  {http://arxiv.org/abs/0912.5317} {arXiv:0912.5317 [astro-ph.CO]} \BibitemShut
  {NoStop}%
\bibitem [{\citenamefont {Ivanov}\ \emph {et~al.}(1994)\citenamefont {Ivanov},
  \citenamefont {Naselsky},\ and\ \citenamefont {Novikov}}]{Ivanov:1994pa}%
  \BibitemOpen
  \bibfield  {author} {\bibinfo {author} {\bibfnamefont {P.}~\bibnamefont
  {Ivanov}}, \bibinfo {author} {\bibfnamefont {P.}~\bibnamefont {Naselsky}}, \
  and\ \bibinfo {author} {\bibfnamefont {I.}~\bibnamefont {Novikov}},\ }\href
  {\doibase 10.1103/PhysRevD.50.7173} {\bibfield  {journal} {\bibinfo
  {journal} {Phys. Rev.}\ }\textbf {\bibinfo {volume} {D50}},\ \bibinfo {pages}
  {7173} (\bibinfo {year} {1994})}\BibitemShut {NoStop}%
%%CITATION = PHRVA,D50,7173;%%
\bibitem [{\citenamefont {Saito}\ \emph {et~al.}(2008)\citenamefont {Saito},
  \citenamefont {Yokoyama},\ and\ \citenamefont {Nagata}}]{Saito:2008em}%
  \BibitemOpen
  \bibfield  {author} {\bibinfo {author} {\bibfnamefont {R.}~\bibnamefont
  {Saito}}, \bibinfo {author} {\bibfnamefont {J.}~\bibnamefont {Yokoyama}}, \
  and\ \bibinfo {author} {\bibfnamefont {R.}~\bibnamefont {Nagata}},\ }\href
  {\doibase 10.1088/1475-7516/2008/06/024} {\bibfield  {journal} {\bibinfo
  {journal} {J. Cosmol. Astropart. Phys.}\ }\textbf {\bibinfo {volume}
  {0806}},\ \bibinfo {pages} {024} (\bibinfo {year} {2008})},\ \Eprint
  {http://arxiv.org/abs/0804.3470} {arXiv:0804.3470 [astro-ph]} \BibitemShut
  {NoStop}%
%%CITATION = 0804.3470;%%
\bibitem [{\citenamefont {Motohashi}\ and\ \citenamefont
  {Hu}(2017)}]{Motohashi:2017kbs}%
  \BibitemOpen
  \bibfield  {author} {\bibinfo {author} {\bibfnamefont {H.}~\bibnamefont
  {Motohashi}}\ and\ \bibinfo {author} {\bibfnamefont {W.}~\bibnamefont {Hu}},\
  }\href {\doibase 10.1103/PhysRevD.96.063503} {\bibfield  {journal} {\bibinfo
  {journal} {Phys. Rev.}\ }\textbf {\bibinfo {volume} {D96}},\ \bibinfo {pages}
  {063503} (\bibinfo {year} {2017})},\ \Eprint
  {http://arxiv.org/abs/1706.06784} {arXiv:1706.06784 [astro-ph.CO]}
  \BibitemShut {NoStop}%
%%CITATION = ARXIV:1706.06784;%%
\bibitem [{\citenamefont {Cai}\ \emph {et~al.}(2019)\citenamefont {Cai},
  \citenamefont {Pi},\ and\ \citenamefont {Sasaki}}]{Cai:2018dig}%
  \BibitemOpen
  \bibfield  {author} {\bibinfo {author} {\bibfnamefont {R.-g.}\ \bibnamefont
  {Cai}}, \bibinfo {author} {\bibfnamefont {S.}~\bibnamefont {Pi}}, \ and\
  \bibinfo {author} {\bibfnamefont {M.}~\bibnamefont {Sasaki}},\ }\href
  {\doibase 10.1103/PhysRevLett.122.201101} {\bibfield  {journal} {\bibinfo
  {journal} {Phys. Rev. Lett.}\ }\textbf {\bibinfo {volume} {122}},\ \bibinfo
  {pages} {201101} (\bibinfo {year} {2019})},\ \Eprint
  {http://arxiv.org/abs/1810.11000} {arXiv:1810.11000 [astro-ph.CO]}
  \BibitemShut {NoStop}%
\bibitem [{\citenamefont {Kinney}(2005)}]{Kinney:2005vj}%
  \BibitemOpen
  \bibfield  {author} {\bibinfo {author} {\bibfnamefont {W.~H.}\ \bibnamefont
  {Kinney}},\ }\href {\doibase 10.1103/PhysRevD.72.023515} {\bibfield
  {journal} {\bibinfo  {journal} {Phys. Rev. D}\ }\textbf {\bibinfo {volume}
  {72}},\ \bibinfo {pages} {023515} (\bibinfo {year} {2005})},\ \Eprint
  {http://arxiv.org/abs/gr-qc/0503017} {arXiv:gr-qc/0503017} \BibitemShut
  {NoStop}%
\bibitem [{\citenamefont {Martin}\ \emph {et~al.}(2013)\citenamefont {Martin},
  \citenamefont {Motohashi},\ and\ \citenamefont {Suyama}}]{Martin:2012pe}%
  \BibitemOpen
  \bibfield  {author} {\bibinfo {author} {\bibfnamefont {J.}~\bibnamefont
  {Martin}}, \bibinfo {author} {\bibfnamefont {H.}~\bibnamefont {Motohashi}}, \
  and\ \bibinfo {author} {\bibfnamefont {T.}~\bibnamefont {Suyama}},\ }\href
  {\doibase 10.1103/PhysRevD.87.023514} {\bibfield  {journal} {\bibinfo
  {journal} {Phys. Rev. D}\ }\textbf {\bibinfo {volume} {87}},\ \bibinfo
  {pages} {023514} (\bibinfo {year} {2013})},\ \Eprint
  {http://arxiv.org/abs/1211.0083} {arXiv:1211.0083 [astro-ph.CO]} \BibitemShut
  {NoStop}%
\bibitem [{\citenamefont {Motohashi}\ \emph {et~al.}(2015)\citenamefont
  {Motohashi}, \citenamefont {Starobinsky},\ and\ \citenamefont
  {Yokoyama}}]{Motohashi:2014ppa}%
  \BibitemOpen
  \bibfield  {author} {\bibinfo {author} {\bibfnamefont {H.}~\bibnamefont
  {Motohashi}}, \bibinfo {author} {\bibfnamefont {A.~A.}\ \bibnamefont
  {Starobinsky}}, \ and\ \bibinfo {author} {\bibfnamefont {J.}~\bibnamefont
  {Yokoyama}},\ }\href {\doibase 10.1088/1475-7516/2015/09/018} {\bibfield
  {journal} {\bibinfo  {journal} {JCAP}\ }\textbf {\bibinfo {volume} {09}},\
  \bibinfo {pages} {018} (\bibinfo {year} {2015})},\ \Eprint
  {http://arxiv.org/abs/1411.5021} {arXiv:1411.5021 [astro-ph.CO]} \BibitemShut
  {NoStop}%
\bibitem [{\citenamefont {Suyama}\ \emph {et~al.}(2021)\citenamefont {Suyama},
  \citenamefont {Tada},\ and\ \citenamefont {Yamaguchi}}]{Suyama:2021adn}%
  \BibitemOpen
  \bibfield  {author} {\bibinfo {author} {\bibfnamefont {T.}~\bibnamefont
  {Suyama}}, \bibinfo {author} {\bibfnamefont {Y.}~\bibnamefont {Tada}}, \ and\
  \bibinfo {author} {\bibfnamefont {M.}~\bibnamefont {Yamaguchi}},\ }\href
  {\doibase 10.1093/ptep/ptab063} {\bibfield  {journal} {\bibinfo  {journal}
  {PTEP}\ }\textbf {\bibinfo {volume} {2021}},\ \bibinfo {pages} {073E02}
  (\bibinfo {year} {2021})},\ \Eprint {http://arxiv.org/abs/2101.10682}
  {arXiv:2101.10682 [hep-th]} \BibitemShut {NoStop}%
\bibitem [{\citenamefont {Firouzjahi}\ \emph {et~al.}(2019)\citenamefont
  {Firouzjahi}, \citenamefont {Nassiri-Rad},\ and\ \citenamefont
  {Noorbala}}]{Firouzjahi:2018vet}%
  \BibitemOpen
  \bibfield  {author} {\bibinfo {author} {\bibfnamefont {H.}~\bibnamefont
  {Firouzjahi}}, \bibinfo {author} {\bibfnamefont {A.}~\bibnamefont
  {Nassiri-Rad}}, \ and\ \bibinfo {author} {\bibfnamefont {M.}~\bibnamefont
  {Noorbala}},\ }\href {\doibase 10.1088/1475-7516/2019/01/040} {\bibfield
  {journal} {\bibinfo  {journal} {JCAP}\ }\textbf {\bibinfo {volume} {01}},\
  \bibinfo {pages} {040} (\bibinfo {year} {2019})},\ \Eprint
  {http://arxiv.org/abs/1811.02175} {arXiv:1811.02175 [hep-th]} \BibitemShut
  {NoStop}%
\bibitem [{\citenamefont {Namjoo}\ \emph {et~al.}(2013)\citenamefont {Namjoo},
  \citenamefont {Firouzjahi},\ and\ \citenamefont {Sasaki}}]{Namjoo:2012aa}%
  \BibitemOpen
  \bibfield  {author} {\bibinfo {author} {\bibfnamefont {M.~H.}\ \bibnamefont
  {Namjoo}}, \bibinfo {author} {\bibfnamefont {H.}~\bibnamefont {Firouzjahi}},
  \ and\ \bibinfo {author} {\bibfnamefont {M.}~\bibnamefont {Sasaki}},\ }\href
  {\doibase 10.1209/0295-5075/101/39001} {\bibfield  {journal} {\bibinfo
  {journal} {EPL}\ }\textbf {\bibinfo {volume} {101}},\ \bibinfo {pages}
  {39001} (\bibinfo {year} {2013})},\ \Eprint {http://arxiv.org/abs/1210.3692}
  {arXiv:1210.3692 [astro-ph.CO]} \BibitemShut {NoStop}%
\end{thebibliography}%

\end{document}